\documentclass[acmsmall]{acmart}

\setcopyright{cc}
\setcctype{by}
\acmJournal{POMACS}
\acmYear{2026} \acmVolume{10} \acmNumber{3} \acmArticle{66}
\acmMonth{12} 
\acmDOI{10.1145/3857369}

\usepackage{colortbl} 
\usepackage{graphicx}

\usepackage{arydshln} 
\usepackage{makecell}
\usepackage{array}
\usepackage{enumitem}
\usepackage{subcaption}
\usepackage{comment}
\usepackage{booktabs}
\usepackage{multirow}
\usepackage{graphicx}
\usepackage{booktabs}
\usepackage{multirow}
\usepackage{algorithm}
\usepackage{algorithmic}

\usepackage[english]{babel}

\usepackage{fancyhdr} 
\usepackage{float}
\usepackage{subcaption}

\usepackage{pifont} 
\usepackage{bbding} 

\usepackage{tcolorbox}
\tcbuselibrary{skins,breakable}

\newtcolorbox{strategyhead}{
  enhanced,
  colback=blue!4!white,    
  colframe=blue!55!black,  
  boxrule=0.6pt,
  arc=2pt,
  left=4pt, right=4pt, top=0pt, bottom=0pt,
}

\newtcolorbox{flawhead}{
  enhanced,
  colback=red!4!white,
  colframe=red!55!black,
  boxrule=0.6pt,
  arc=2pt,
  left=4pt, right=4pt, top=0pt, bottom=0pt,
}

\newtcolorbox{breakdownhead}{
  enhanced,
  colback=teal!5!white,
  colframe=teal!70!black,
  boxrule=0.6pt, arc=2pt,
  left=4pt, right=4pt, top=0pt, bottom=0pt,
}

\usepackage{tikz}

\definecolor{ppoifail}{RGB}{200,40,40}
\definecolor{ppoipass}{RGB}{40,140,60}

\usepackage{url}

\usepackage{hyperref}
\newenvironment{packeditemize}{
	\begin{list}{$\bullet$}{
			\setlength{\labelwidth}{4pt}
			\setlength{\itemsep}{0pt}
			\setlength{\leftmargin}{\labelwidth}
			\addtolength{\leftmargin}{\labelsep}
			\setlength{\parindent}{0pt}
			\setlength{\listparindent}{\parindent}
			\setlength{\parsep}{0pt}
			\setlength{\topsep}{1pt}}}{\end{list}}

\usepackage{xcolor}
\usepackage{xspace} 

\definecolor{empirical}{RGB}{160, 32, 240}

\newcommand{\modified}[1]{{\color{black}#1}}

\AtBeginDocument{%
  }


\usepackage{acro}
\acsetup{single}

\DeclareAcronym{EOA}{
  short = EOA,
  long  =  Externally-Owned Account,
}

\DeclareAcronym{CEX}{
  short = CEX,
  long  = Centralized Exchange,
}

\DeclareAcronym{OFAC}{
  short = OFAC,
  long  = Office of Foreign Assets Control,
}

\DeclareAcronym{TC}{
  short = TC,
  long  = Tornado Cash,
}
\newcommand{\TC}{\ac{TC}\xspace}

\DeclareAcronym{RG}{
  short = RG,
  long  = Railgun,
}
\newcommand{\RG}{\ac{RG}\xspace}

\begin{document}


\title[How Attackers Launder Funds through Mixers: Characterization and Detection on Ethereum]{How Attackers Launder Funds through Mixers:\\ Characterization and Detection on Ethereum}


\author{Qishuang Fu}
\orcid{0009-0000-6964-6641}
\authornote{Corresponding author: Qishuang Fu (qishuang.fu@monash.edu)}
\email{qishuang.fu@monash.edu}

\affiliation{%
  \institution{Monash University}
   \city{Melbourne}
  \country{Australia}
 }

\author{Hang Zheng}
\orcid{0009-0004-7165-5984}
\email{hzhe0079@student.monash.edu}
\affiliation{%
  \institution{Monash University}
   \city{Melbourne}
  \country{Australia}}

\author{Xihan Xiong}
\orcid{0009-0002-1604-0114}
\email{xihan.xiong20@imperial.ac.uk}
\affiliation{%
  \institution{Imperial College London}
  \city{London}
  \country{UK}}

\author{Joseph K. Liu}
\orcid{0000-0001-6656-6240}
\email{joseph.liu@monash.edu}
\affiliation{%
  \institution{Monash University}
   \city{Melbourne}
  \country{Australia}}

\author{Yixin Liu}
\orcid{0000-0002-4309-5076}
\email{yixin.liu@griffith.edu.au}
\affiliation{%
\institution{Griffith University}
\city{Brisbane}
\country{Australia}}

\author{Shirui Pan}
\orcid{0000-0003-0794-527X}
\email{s.pan@griffith.edu.au}
\affiliation{%
  \institution{Griffith University}
  \city{Gold Coast}
  \country{Australia}}

\author{Qin Wang}
\orcid{0000-0003-2264-749X}
\email{qinwangtech@gmail.com}
\affiliation{%
  \institution{CSIRO}
  \city{Sydney}
  \country{Australia}}

\author{Weiqing Wang}
\orcid{0000-0002-9578-819X}
\email{teresa.wang@monash.edu}
\affiliation{%
  \institution{Monash University}
   \city{Melbourne}
  \country{Australia}}

\author{Zhipeng Wang}
\orcid{0000-0003-1089-1583}
\email{zhipeng.wang@manchester.ac.uk}
\affiliation{%
  \institution{The University of Manchester}
  \city{Manchester}
  \country{UK}}

\author{Tsz Hon Yuen}
\orcid{0000-0002-0629-6792}
\email{john.tszhonyuen@monash.edu}
\affiliation{%
  \institution{Monash University}
   \city{Melbourne}
  \country{Australia}}

\renewcommand{\shortauthors}{Qishuang Fu et al.}

\begin{abstract}

Mixers protect privacy by concealing deposit--withdrawal links, but are also abused to launder illicit funds. Existing anti-money laundering studies do not specifically target mixer laundering, while mixer research focuses on deanonymization rather than identifying laundering-related transactions. Public reports remain fragmented, leaving no public case-level dataset for systematic measurement and detection. To fill this void, this paper presents the first comprehensive study of mixer laundering on Ethereum. We first construct \textsc{MixLaunder}, the first public case-level dataset of mixer laundering. It covers 27 cases involving Tornado Cash and Railgun from 2020 to 2025 and labels 9,300 laundering-related transactions with case identities and observable upstream and downstream fund flows, including deposits totaling approximately \$1.1 billion. By comparing these transactions with background mixer usage, we identify five common strategies, showing that laundering evidence spans complementary behavioral and fund-flow contexts, while same-case activity is locally tight but weakly connected across bursts. Our analysis further reveals coverage gaps in mixer-side risk screening and representative deanonymization heuristics. Guided by these findings, we develop \textsc{MixGuard}, which combines tri-view representation learning with two-stage grouping for transaction-level detection and case-aware grouping. Under strict case-level holdout evaluation, \textsc{MixGuard} outperforms representative baselines, achieving 95.76\% detection precision and 98.40\% group purity, while its top ten groups cover 95.95\% of each case's transactions on average. Applied to Ethereum data from January 2026, \textsc{MixGuard} identifies and validates four incidents involving over \$15 million, which we responsibly disclosed to Etherscan and Tornado Cash. Our work aims to serve as a valuable reference for combating mixer laundering, advancing on-chain risk detection, and preserving legitimate privacy.

\end{abstract}


\maketitle

\section{Introduction}
\label{sec-intro}

Recent years have witnessed substantial cryptocurrency losses from protocol exploits, phishing thefts, and cross-chain bridge attacks across the Ethereum ecosystem~\cite{wu2023towards,huo2025shedding,he2023txphishscope}. For attackers, stealing assets is only the first step. 
To further transfer, exchange, or cash out the proceeds, they seek to obscure the on-chain links between the source of the funds and their subsequent destinations. Attackers therefore repeatedly route stolen funds through mixers such as \TC~\cite{youn2023empirical,wang2023zkmixer} and \RG~\cite{railgun-ppoi}. 
By using cryptographic mechanisms such as zero-knowledge proofs, mixers weaken publicly observable links between deposits and withdrawals, enabling funds to be withdrawn to fresh addresses without a visible link to specific deposits~\cite{glaeser2022foundations,wang2023zkmixer,railgun-ppoi}. 
We refer to this post-attack use of mixers, together with its pre-mixing preparation and post-mixing disposal, as \textbf{\textit{mixer laundering}}. 
Although mixers also serve legitimate privacy needs, public investigations have repeatedly documented their use in laundering campaigns associated with Lazarus and other threat actors~\cite{wang2023zkmixer,tayvano_lazarus_bluenoroff_hacks,zachxbt_lazarus_laundered_200m}. 
The practical investigative task is therefore to distinguish laundering transactions from background mixer activity and further determine which suspicious transactions belong to the same attack case.

However, available data and research remain insufficient to support this task. Public reports typically disclose only key addresses, transaction hashes, or fragments of fund flows, while existing Ethereum laundering datasets do not provide a case-level view of mixer laundering~\cite{wu2023towards}. 
Existing blockchain anti-money laundering~(AML) studies~\cite{weber2019anti,lin2024denseflow,li2025multi} do not specifically address laundering across the privacy boundary introduced by mixers.
Meanwhile, prior mixer studies primarily investigate privacy properties~\cite{glaeser2022foundations,youn2023empirical,wang2023zkmixer} or deanonymization, typically seeking candidate links between deposits and withdrawals~\cite{du2024breaking,Wang2025improving,tc-clustering}. Yet a linked deposit--withdrawal pair does not by itself establish laundering, and real laundering activity may leave no reliable pairwise link. Deanonymization asks which withdrawal may correspond to a deposit, whereas laundering investigation must determine which transactions are associated with illicit funds and which suspicious transactions belong to the same case. Consequently, the community lacks both a systematic measurement of real-world mixer laundering and a method that jointly supports transaction-level laundering detection and case-aware grouping.

Addressing this gap cannot rely on recovering exact deposit--withdrawal links. The key question is whether mixer laundering exhibits common observable characteristics that distinguish it from background mixer usage, and whether these characteristics can support transaction-level detection and case-aware grouping without compromising mixer privacy.


\noindent\textbf{Our Work.}
To answer these questions, we follow a measurement-first approach that characterizes mixer laundering and then translates the findings into detection. We begin by constructing \textsc{MixLaunder}, the first public case-level dataset of mixer laundering on Ethereum. \textsc{MixLaunder} covers 27 real-world attack cases involving Tornado Cash and Railgun between 2020 and 2025, providing laundering labels, case labels, and observable upstream and downstream fund flows for their mixer transactions. It contains 9,300 laundering-related mixer transactions, through which approximately \$1.1 billion in illicit funds were deposited into mixers. Using \textsc{MixLaunder}, we compare real-world laundering transactions with background mixer usage and identify five common strategies spanning fund preparation, mixer use, post-withdrawal restructuring, and intra-case coordination. These strategies show that no single perspective fully characterizes laundering transactions, while relying only on strong local relationships leaves same-case activity fragmented. We further evaluate existing countermeasures and identify gaps in mixer-side risk screening. Even the broadest of three representative deanonymization heuristics covers only 4 of the 27 cases.

Guided by these findings, we design \textsc{MixGuard}, a unified framework for transaction-level laundering detection and case-aware grouping. Its tri-view encoder integrates address, fund-flow path, and local topology information, while its two-stage grouping forms reliable local cores before merging related cores through weaker ties. Under strict case-level holdout evaluation, \textsc{MixGuard} outperforms representative baselines, achieving 95.76\% detection precision, 98.40\% group purity, and 95.95\% average top-ten case coverage. Case studies show that the recovered groups capture laundering stages spanning multiple addresses and time windows. Applied to Ethereum data from January 2026, it identifies and validates four mixer-laundering incidents involving more than \$15 million, which we responsibly disclosed to Etherscan and Tornado Cash. 

\noindent\textbf{Contributions.} In summary, we make the following contributions:
\begin{packeditemize}

    \item \textbf{First Public Case-Level Dataset of Mixer Laundering.} 
We construct and release \textsc{MixLaunder}, providing laundering labels, case labels, and observable upstream and downstream fund flows for mixer transactions from real-world attack cases.

    \item \textbf{Systematic Measurement Grounded in Real-World Cases.} 
We identify five common strategies of mixer laundering and evaluate existing countermeasures, revealing coverage limitations in mixer-side risk screening and address-level deanonymization heuristics.

    \item \textbf{Measurement-Guided Detection and Grouping Framework.} 
We propose \textsc{MixGuard}, which combines tri-view representation learning with two-stage grouping to support transaction-level laundering detection and case-aware grouping. We validate its effectiveness through strict evaluation on unseen cases and real-world deployment. Our dataset and code are publicly available at: \url{https://anonymous.4open.science/r/MixGuard-62A3}. 
\end{packeditemize}

\begin{figure}[t]
    \centering
                \begin{subfigure}[b]{0.4\linewidth}
        \centering
        \includegraphics[width=\linewidth]{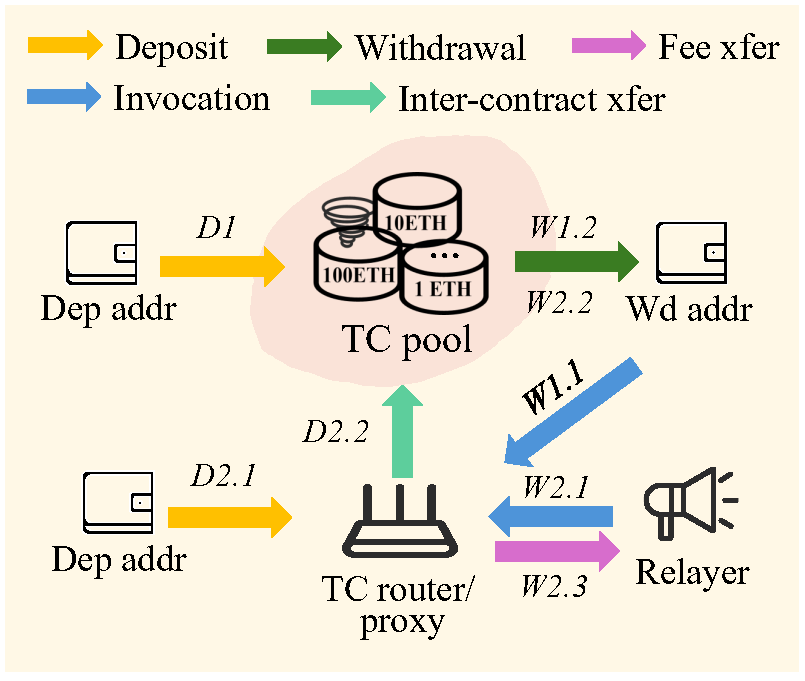}
        \caption{Tornado Cash}
        \label{fig:tc-tx}
    \end{subfigure}
        \hspace{0.02\linewidth}
        \begin{subfigure}[b]{0.4\linewidth}
        \centering
        \includegraphics[width=\linewidth]{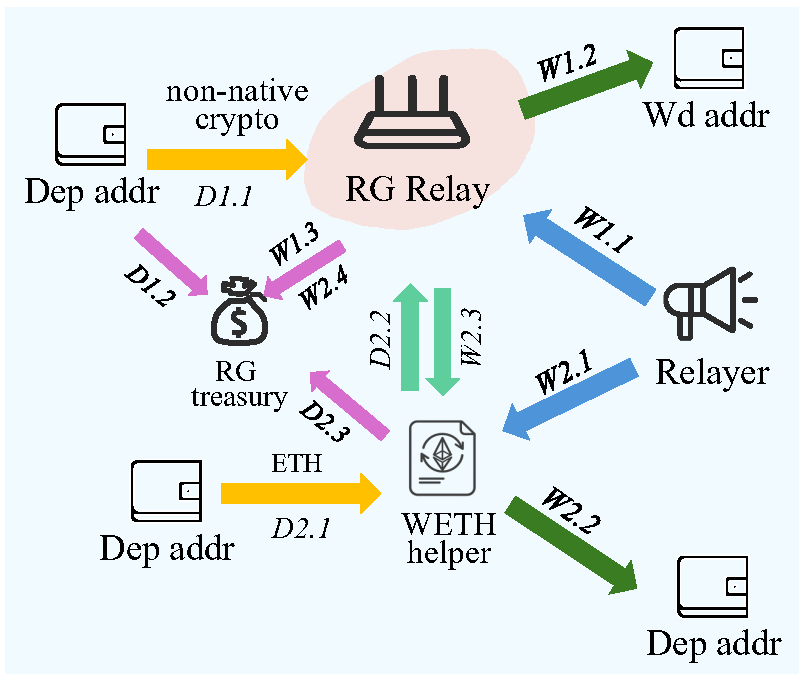}
        \caption{Railgun}
        \label{fig:rg-tx}
            \end{subfigure}

    \caption{Typical deposit and withdrawal transaction structures of the two mixers.}
    \label{fig:mixer-tx}
\end{figure}

\section{Preliminaries}
\label{sec-bck}

We start by briefly introducing the target mixers~(\S~\ref{subsec-mixer}), and then describe the mixer laundering workflow~(\S\ref{subsec-workflow}). 

\begin{figure}[t]
    \centering
    \includegraphics[width=0.6\linewidth]{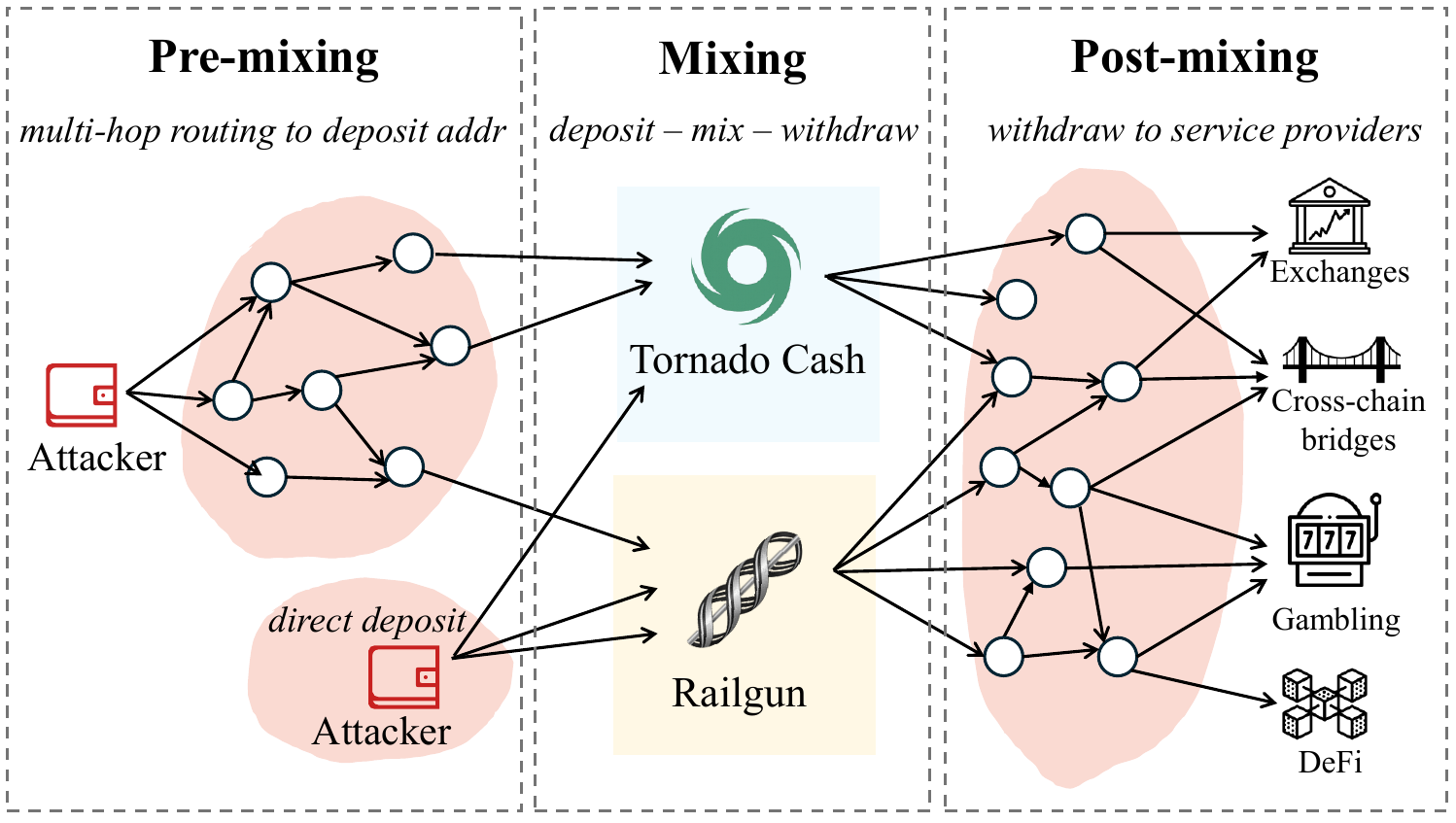}
    \caption{Mixer laundering workflow. Funds pass through three stages: \textit{pre-mixing preparation}, \textit{mixing}, and \textit{post-mixing disposal}. Deposits and withdrawals are visible on
    chain, while their exact correspondence is hidden by the mixer.}
    \label{fig:workflow}
\end{figure}

\subsection{Target Mixers}
\label{subsec-mixer}

We focus on \TC and \RG on Ethereum for two reasons. First, they are dominant
privacy mixers on Ethereum. In 2025, \TC and \RG together accounted for about
$95\%$ of mixer transactions on Ethereum~\cite{cambridge-mixers-2026} and both have been reported in real mixer laundering
cases~\cite{fbi-tornado-2023,elliptic-harmony-railgun-2023}. Second, they
represent different on-chain privacy mechanisms: \TC is based on
fixed-denomination pools~\cite{youn2023empirical}, while \RG is based on \modified{private balance and transfers that conceal transaction parties and amounts, with funds
withdrawn to public addresses~\cite{railgun-privacy-system}. It also supports Private Proofs of
Innocence (POI)~\cite{railgun-ppoi}, which uses zero-knowledge proofs to demonstrate that funds are
not associated with addresses or transactions on specified risk lists}.

Figure~\ref{fig:mixer-tx} shows the typical deposit and
withdrawal structures. In \TC~(Figure~\ref{fig:tc-tx}), deposits either go directly from the deposit
address to a fixed-denomination pool~(\emph{D1}), or enter the pool through a
router/proxy~(\emph{D2.1--D2.2}). Withdrawals are invoked by the user or a relayer
(\emph{W1.1/W2.1}), after which the pool releases the fixed-denomination amount to the
withdrawal address~(\emph{W1.2/W2.2}). In \RG~(Figure~\ref{fig:rg-tx}), non-native assets are deposited through
the relay~(\emph{D1.1}), while ETH deposits pass through the WETH helper before
entering the relay~(\emph{D2.1--D2.2}). Withdrawals release funds from the relay or
helper to withdrawal addresses~(\emph{W1.2/W2.2}).



These structures create a measurement issue: a single transaction may include
router calls, relayer payments, helper-contract interactions, and token transfers,
only one of which corresponds to the actual mixer deposit or withdrawal. We
therefore extract standardized deposit and withdrawal records from raw
transactions before constructing labels and fund-flow context (§\ref{subsec-mixer-extraction}).

\subsection{Mixer Laundering Workflow}
\label{subsec-workflow}

Mixer laundering usually involves a sequence of fund movements around the
mixer, rather than a single deposit or withdrawal. Illicit funds are first moved
toward the mixer, then deposited and withdrawn through the mixer, and finally
moved again after withdrawal. Figure~\ref{fig:workflow} shows this three-stage
workflow.

\begin{packeditemize}
  \item \textit{\underline{Pre-mixing preparation.}}
   Illicit funds move from attack-related addresses through intermediate
  transactions, and eventually reach the address that initiates the mixer
  deposit. This stage happens before the mixer and is publicly visible on chain.

  \item \textit{\underline{Mixing.}}
  Funds enter the mixer through observable deposit transactions and later leave
through observable withdrawal transactions. However, the mixer's privacy
mechanism~\cite{glaeser2022foundations} hides which withdrawal corresponds to
which deposit. As a result, an on-chain observer can see funds entering and
leaving the mixer, but cannot directly link a specific withdrawal to a specific
deposit.

  \item \textit{\underline{Post-mixing disposal.}}
  After leaving the mixer, funds continue through transfers, swaps, splits,
  aggregations, or services such as exchanges and bridges. This stage happens
  after the mixer, and the related fund flows are again visible on chain.
\end{packeditemize}


We model a mixer laundering case as  $\mathcal{C} =
  \bigl(G_{\mathit{up}},\mathcal{T}_{\mathit{dep}},
  \mathcal{T}_{\mathit{wd}},G_{\mathit{dn}}\bigr)$,
where $\mathcal{T}_{\mathit{dep}}$ and $\mathcal{T}_{\mathit{wd}}$
are the sets of mixer deposit and withdrawal transactions in the
case, and $G_{\mathit{up}}$ and $G_{\mathit{dn}}$ are the upstream
and downstream fund-flow graphs that precede the deposits and
follow the withdrawals, respectively.

\section{Study Design}
\label{sec-study-design}

In this work, we perform a progressive study to have a comprehensive understanding of mixer laundering on Ethereum. Specifically, we aim to answer the following research questions~(RQs):

\vspace{2pt}
\noindent\textbf{RQ1: How can we build a case-level dataset for mixer laundering?}
Public reports often disclose key addresses, transaction or fund-flow snippets, but rarely provide a case-level view of the full observable laundering workflow: how funds reach the mixer, which deposits and withdrawals are laundering-related, and where funds move after withdrawal. 
Existing Ethereum laundering datasets~\cite{wu2023towards} do not explicitly label mixer-laundering deposits and withdrawals, link them to specific cases, or provide the corresponding upstream and downstream fund flows.
Without such an end-to-end case-level dataset, it is difficult to measure mixer laundering as a complete workflow.

\vspace{2pt}
\noindent\textbf{RQ2: What are the measurable characteristics of mixer laundering?}
Although public reports have shown that attackers abuse mixers to launder funds, the community still lacks a systematic measurement of the common patterns shared by real mixer laundering cases. 
In particular, it remains unclear how attackers prepare funds before mixing, schedule deposits and withdrawals, reshape funds after withdrawal, and coordinate multiple addresses. 
Figuring out these characteristics can help understand how mixer laundering differs from background mixer usage.

\vspace{2pt}
\noindent\textbf{RQ3: How can the measured characteristics support laundering detection and case-level group recovery?}
After characterizing mixer laundering in RQ2, a natural next step is to ask whether these characteristics can support practical investigation. Investigators need to identify laundering deposits and withdrawals from background mixer usage, and further recover which transactions belong to the same case.
This is challenging because mixers hide exact deposit--withdrawal links, and laundering actors intentionally fragment same-case activity across addresses, time windows, and fund-flow paths. Answering this question helps determine whether mixer laundering can be detected from observable on-chain behavior without breaking mixer privacy.

\begin{figure}[t]
\centering
\includegraphics[width=\textwidth]{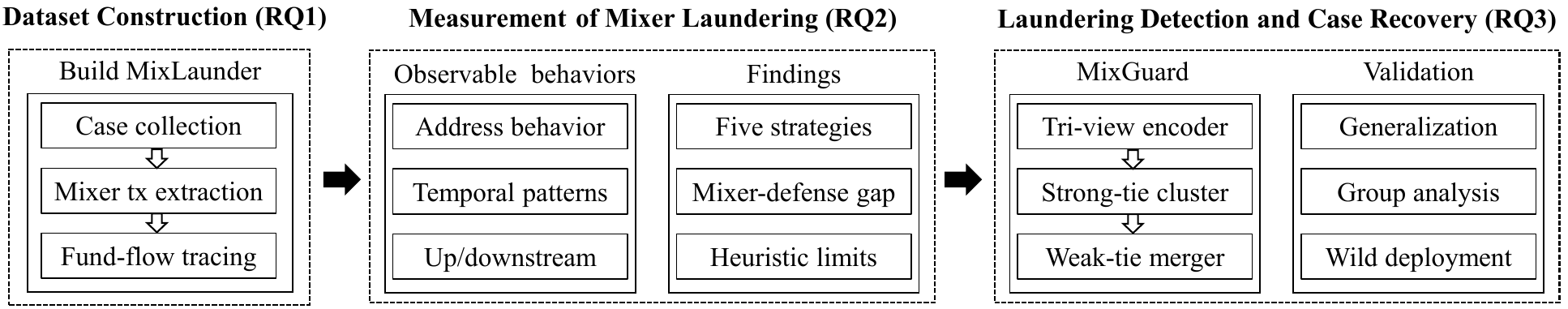}
\caption{An Overview of Our Work.}
\label{fig:overview-0703}
\end{figure}

Figure~\ref{fig:overview-0703} shows the overall workflow of our study. We begin by constructing a ground truth dataset  \textsc{MixLaunder} from the public reports and on-chain transaction records, comprising 27 real-world laundering cases with labeled mixer deposits, withdrawals, and upstream/downstream fund flows. Using this dataset, we measure the laundering workflow before, through, and after mixers, and summarize five common strategies that distinguish laundering from background mixer usage. This analysis also reveals why existing mixer defenses and deanonymization heuristics provide limited coverage. Guided by these measured strategies, we design \textsc{MixGuard} for suspicious
mixer-transaction detection and case-aware grouping. Finally, we evaluate whether \textsc{MixGuard} generalizes to unseen attacks and deploy it on recent Ethereum data to assess its practical value.

\section{Dataset Construction}
\label{sec-dataset}


We build \textsc{MixLaunder}, a dataset for mixer laundering
characterization and detection. 
\modified{
Public reports provide case attributions, mixer deposit and withdrawal clues, and existing fund-flow information. Based on these materials, we extract standardized deposit and withdrawal records and use on-chain data to reconstruct the upstream and downstream fund flows outside the mixers.
}
Its construction has three steps:
case collection (\S\ref{sub-case}), mixer transaction extraction
(\S\ref{subsec-mixer-extraction}), and fund-flow tracing
(\S\ref{subsec-tracing}). We then summarize the resulting \textsc{MixLaunder} laundering and
\textsc{background} datasets (\S\ref{subsec-dataset-overview}). 

\subsection{Case Collection}
\label{sub-case}

We collect mixer-laundering case evidence from public on-chain investigation
reports, including attack-incident repositories~\cite{tayvano_lazarus_bluenoroff_hacks} and cases curated by
on-chain investigators~\cite{zachxbt_lazarus_laundered_200m}. 
\modified{We select these public reports because they document mixer laundering
activity associated with attack incidents and disclose transaction hashes, deposit and withdrawal addresses, or fund-flow graphs. This information provides verifiable on-chain evidence of case-related mixer activity and a traceable starting point for collecting the corresponding deposit and withdrawal transactions and reconstructing their upstream and downstream fund flows.}

\modified{
For each case, we use the transaction hashes or deposit and withdrawal
addresses provided in the public reports to retrieve on-chain records
and verify the corresponding \TC or \RG deposits and withdrawals.
We adopt the case attributions provided by the public investigations
and describe their supporting on-chain evidence and reasoning in
Appendix~\ref{app:case-clues}.
}
Based on the \modified{public evidence and our on-chain checks}, a case is included only if:
\emph{(i)} it is a publicly disclosed real attack;
\emph{(ii)} public materials state that the case used \TC or \RG for laundering;
\emph{(iii)} it provides verifiable evidence that can yield at least one deposit
record and one withdrawal record in the next step.
Cases that fail these conditions and cannot be verified on chain are excluded.

\subsection{Mixer Transaction Extraction}
\label{subsec-mixer-extraction}
After collecting case-level laundering evidence, we convert candidate evidence into
standardized mixer deposit and withdrawal records. We first collect candidate raw
Ethereum transactions. For transaction-hash evidence, we fetch the transaction details from the Etherscan API~\cite{etherscan-api}
and check whether the transaction matches the deposit or withdrawal structures
in {\S\ref{subsec-mixer}}. For address-level clues, we collect \TC and \RG
deposit and withdrawal candidates associated with the reported addresses within the
case-related time window. We keep only candidates that match the public case
description. These transactions are labeled as mixer laundering transactions for the case.


Next, we extract a standardized deposit or withdrawal transfer from each mixer-related transaction. Since a single on-chain transaction may contain multiple transfers, we retain the transfer that captures the actual fund movement into or out of the mixer. Detailed extraction procedures are provided in Appendix~\ref{app:mixer-extraction}. Each extracted record includes the transaction hash, timestamp, sender address, receiver address, asset type, and transfer amount, and serves as a seed for subsequent fund-flow tracing.

\subsection{Fund-Flow Tracing}
\label{subsec-tracing}

After extracting mixer deposit and withdrawal records, we trace how
these funds move before and after the mixer. For each deposit, we trace backward
from the deposited funds to identify the transactions they passed through before
entering the mixer. For each withdrawal, we trace forward from the withdrawn
funds to identify the transactions they flow through after leaving the mixer.
\modified{
We apply consistent tracing rules to on-chain records to reconstruct transactions upstream of mixer deposits and downstream of withdrawals,
together with their attributable amounts. This process extends the case clues provided in public reports into structured fund-flow data for analysis.
}

Our tracing is based on fund attribution, rather than simply expanding all
transactions of the address. 
At each hop, we keep only temporally consistent transactions that can
carry the current traced asset and amount, and record only the portion
attributable to the current mixer deposit or withdrawal record.
When funds pass through a DEX, router, or liquidity
pool, we parse event logs and internal transfers to connect the input asset
before the swap with the output asset after the swap, allowing tracing to continue across swaps. Tracing stops when no attributable amount remains, no related
transaction is found, or funds enter a labeled terminal service such as an
exchange. We use Moralis labels~\cite{moralis-data-api} to identify
these services. Detailed tracing procedures and pseudocode are provided in
Appendix~\ref{app:tracing}.

\subsection{Dataset Overview}
\label{subsec-dataset-overview}




Following the above process, we build \textsc{MixLaunder}, a mixer laundering
dataset covering 27 real-world attack cases from 2020 to 2025. It labels
whether each \TC or \RG deposit and withdrawal is laundering-related and which
case it belongs to. 
Following the same-day background construction used in recent blockchain datasets such as Real-CATS~\cite{shi2025real}, we construct background mixer transactions from \modified{the same UTC calendar days} when each laundering case used the mixer to deposit or withdraw funds, after removing all labeled laundering transactions. \modified{
Although the background transactions may contain unlabeled laundering transactions, this kind of data construction reduces temporal confounding by ensuring that laundering and background transactions are observed under comparable on-chain conditions.} It therefore provides a realistic reference for comparing laundering activity with background mixer usage during the same operational period.

Table~\ref{tab:case-summary} shows detailed statistics of 
our datasets.
\textsc{MixLaunder} contains 9,300 laundering mixer transactions, including
5,349 deposits and 3,951 withdrawals. Starting from these mixer transactions, we
trace 7,916 upstream transactions before deposits and 14,896 downstream
transactions after withdrawals. We apply the same tracing procedure to the
background mixer transactions. The 27 cases vary in value and duration, ranging
from hundreds of thousands of dollars to hundreds of millions of dollars, and
from short mixer use to cases lasting months or longer. For each incident, we also provide a compact case
report with a written summary and a fund-flow diagram, covering the laundering
process and observed tactics. Appendix~\ref{app:case-report} shows one small-case
report example.

\begin{table*}[t]
\centering
\caption{Summary of 27 real-world mixer laundering incidents on Ethereum (2020--2025).
\textbf{Cause}: PKC = Private Key Compromise, SE = Social Engineering,
SC = Smart Contract Exploit, LF = Logic Flaw, API= API Key Abuse.
\textbf{Amount}: total USD value deposited into the mixer.
\textbf{\#Up}/\textbf{\#Dep}/\textbf{\#Wd}/\textbf{\#Down}: number of upstream, deposit, withdrawal, and downstream transactions. \textbf{\#Up}$=0$ indicates that stolen funds are deposited directly into the mixer.
\textbf{Span}: whole-case span in days.}
\label{tab:case-summary}
\setlength{\tabcolsep}{4pt}
\renewcommand{\arraystretch}{1.05}
\resizebox{0.9\textwidth}{!}{%
\begin{tabular}{r l l c l c r r r r r}
\hline
\textbf{\#} & \textbf{Incident} & \textbf{Year} & \textbf{Cause} & \textbf{Amount} & \textbf{Mixer} & \textbf{\#Up} & \textbf{\#Dep} & \textbf{\#Wd} & \textbf{\#Down} & \textbf{Span (d)} \\
\hline
1  & KuCoin                     & 2020 & PKC & \$12.4M   & TC & 2,364 & 287   & 286   & 1,131 & 163.2 \\
2  & HughKarp                   & 2020 & SE  & \$1.6M    & TC & 190   & 26    & 21    & 24    & 182.9 \\
3  & CoinMetro                  & 2020 & PKC & \$5.3M    & TC & 43    & 52    & 45    & 179   & 378.0 \\
\hdashline[1pt/1pt]
4  & BondlyFinance              & 2021 & PKC & \$8.3M    & TC & 1,406 & 109   & 106   & 88    & 411.2 \\
5  & bZx                        & 2021 & PKC & \$45.9M   & TC & 35    & 115   & 115   & 46    & 327.8 \\
6  & LiquidGlobal               & 2021 & PKC & \$25.5M   & TC & 3     & 89    & 55    & 70    & 329.2 \\
7  & Adjacent                   & 2021 & PKC & \$18.6M   & TC & 0     & 71    & 57    & 67    & 20.6 \\
8  & PolyPlay                   & 2021 & PKC & \$1.7M    & TC & 0     & 13    & 5     & 3     & 238.0 \\
\hdashline[1pt/1pt]
9  & Ronin                      & 2022 & PKC & \$449.1M  & TC & 117   & 1,751 & 1,732 & 1,003 & 600.6 \\
10 & Harmony                    & 2022 & PKC & \$94.4M   & TC & 39    & 857   & 180   & 2,387 & 314.6 \\
11 & Deribit                    & 2022 & PKC & \$14.4M   & TC & 1     & 109   & 37    & 1,179 & 366.4 \\
12 & MGNR                       & 2022 & PKC & \$6.7M    & TC & 6     & 45    & 33    & 29    & 374.3 \\
13 & Arthur0x                   & 2022 & PKC & \$1.9M    & TC & 2     & 21    & 7     & 13    & 290.1 \\
\hdashline[1pt/1pt]
14 & OKX                        & 2023 & PKC & \$1.5M    & RG & 14    & 45    & 32    & 95    & 43.5 \\
15 & HTX \& HECO                & 2023 & PKC & \$181.6M  & TC & 1,156 & 606   & 244   & 1,207 & 926.6 \\
16 & Fantom Foundation          & 2023 & PKC & \$3.2M    & TC & 78    & 183   & 160   & 30    & 306.7 \\
17 & SteadeFi \& Coinshift      & 2023 & PKC & \$2.6M    & TC & 3     & 24    & 15    & 102   & 114.9 \\
18 & Hector Network             & 2023 & SC  & \$2.4M    & TC & 70    & 56    & 5     & 8     & 13.2 \\
19 & Shibarium                  & 2023 & SC  & \$1.1M    & TC & 97    & 33    & 33    & 138   & 99.8 \\
20 & Maverick                   & 2023 & PKC & \$10.6M   & TC & 32    & 74    & 55    & 210   & 429.0 \\
\hdashline[1pt/1pt]
21 & WazirX                     & 2024 & PKC & \$152.2M  & TC & 2,140 & 629   & 625   & 6,509 & 1,080.9 \\
22 & ALEX Lab                   & 2024 & PKC & \$7.7M    & Both & 20   & 19    & 21    & 249   & 304.4 \\
23 & ZigCoin                    & 2024 & PKC & \$201.0K  & TC & 0     & 36    & 7     & 12    & 177.5 \\
\hdashline[1pt/1pt]
24 & Garden Finance             & 2025 & LF  & \$15.5M   & RG & 23    & 67    & 50    & 56    & 17.9 \\
25 & Gumi                       & 2025 & SE  & \$2.7M    & TC & 11    & 11    & 4     & 51    & 163.0 \\
26 & JPThor                     & 2025 & PKC & \$241.0K  & TC & 65    & 8     & 8     & 1     & 69.2 \\
27 & VALR                       & 2025 & API & \$145.2K  & TC & 1     & 13    & 13    & 9     & 68.1 \\

\hline
\multicolumn{4}{l}{\textbf{\textsc{MixLaunder} dataset}}
   & \textbf{\$1.1B} & & \textbf{7,916} & \textbf{5,349} & \textbf{3,951} & \textbf{14,896} & \\
\multicolumn{4}{l}{\textbf{Background dataset}}
    & \textbf{\$1.2B} & & \textbf{60,926} & \textbf{9,716} & \textbf{10,996} & \textbf{145,727} &  \\

\hline
\end{tabular}
}
\end{table*}

\section{Characterizing Mixer Laundering}
\label{sec-mixerFeature}


We use \textsc{MixLaunder} to characterize mixer laundering and evaluate existing countermeasures.
We first analyze how laundering actors organize funds before, through, and after mixers and identify five laundering strategies~(\S\ref{sub-stra}). 
We then evaluate protocol-level risk screening and deanonymization heuristics, revealing why they provide limited coverage in real laundering cases~(\S\ref{sub-flaws}).



\subsection{Strategies in Mixer Laundering}
\label{sub-stra}
We derive mixer laundering strategies by comparing laundering and background mixer activity across address profiles, deposit-withdrawal timing, fund flows, and intra-case coordination. These dimensions capture how laundering actors use mixers, prepare and reshape funds around them, and coordinate same-case transactions. We summarize these typical patterns into five strategies.

\begin{figure}[t]
  \centering
  \begin{subfigure}[b]{0.38\linewidth}
    \centering
    \includegraphics[width=\linewidth]{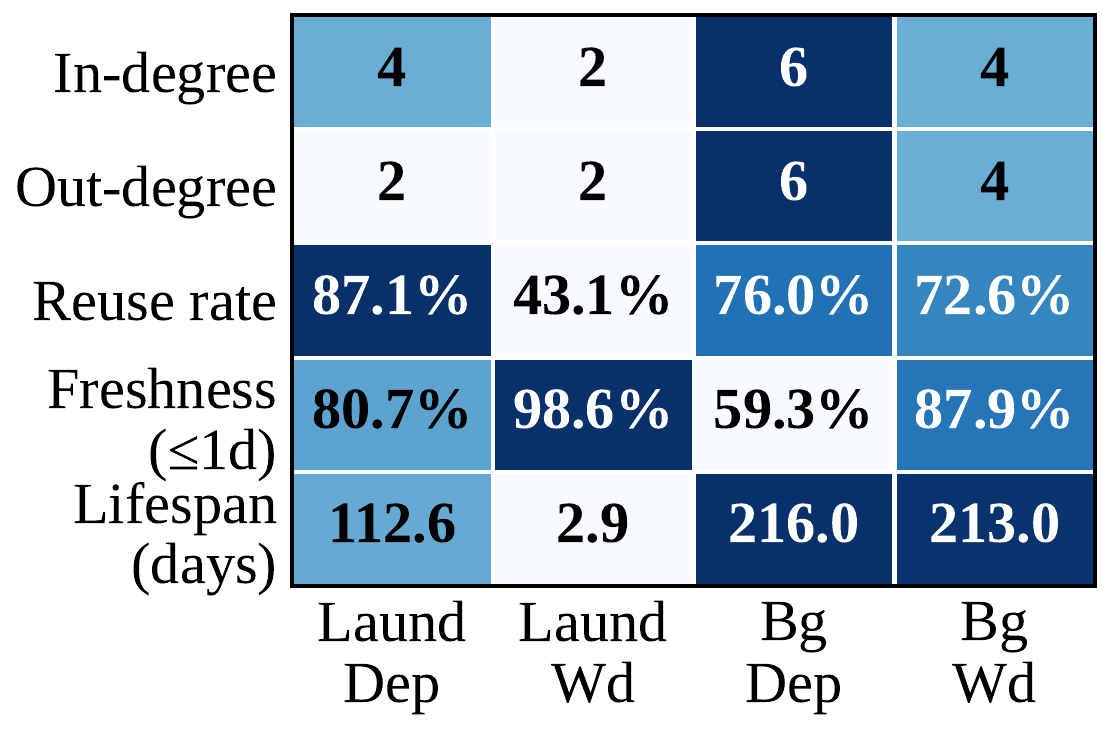}
    \caption{Behavioral profile.}
    \label{fig:address-profile}
  \end{subfigure}
  \hspace{0.02\linewidth}
  \begin{subfigure}[b]{0.36\linewidth}
    \centering
    \includegraphics[width=\linewidth]{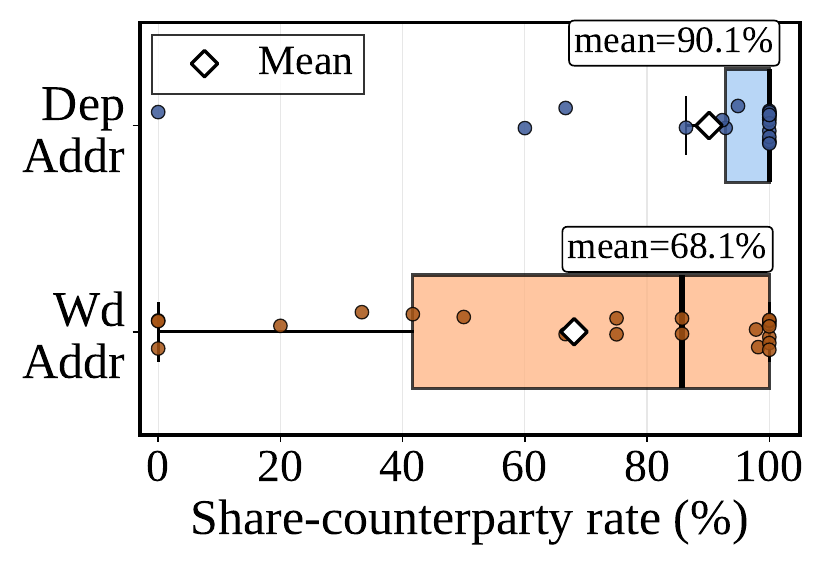}
    \caption{Shared counterparties ratio.}
    \label{fig:within-case-shared-cp}
  \end{subfigure}
  \caption{Behavioral profile and counterparty overlap.}
  \label{fig:profile-and-cp}
\end{figure}



\begin{figure}[t]
    \centering

    \begin{subfigure}[t]{0.44\linewidth}
        \centering
        \includegraphics[width=\linewidth]{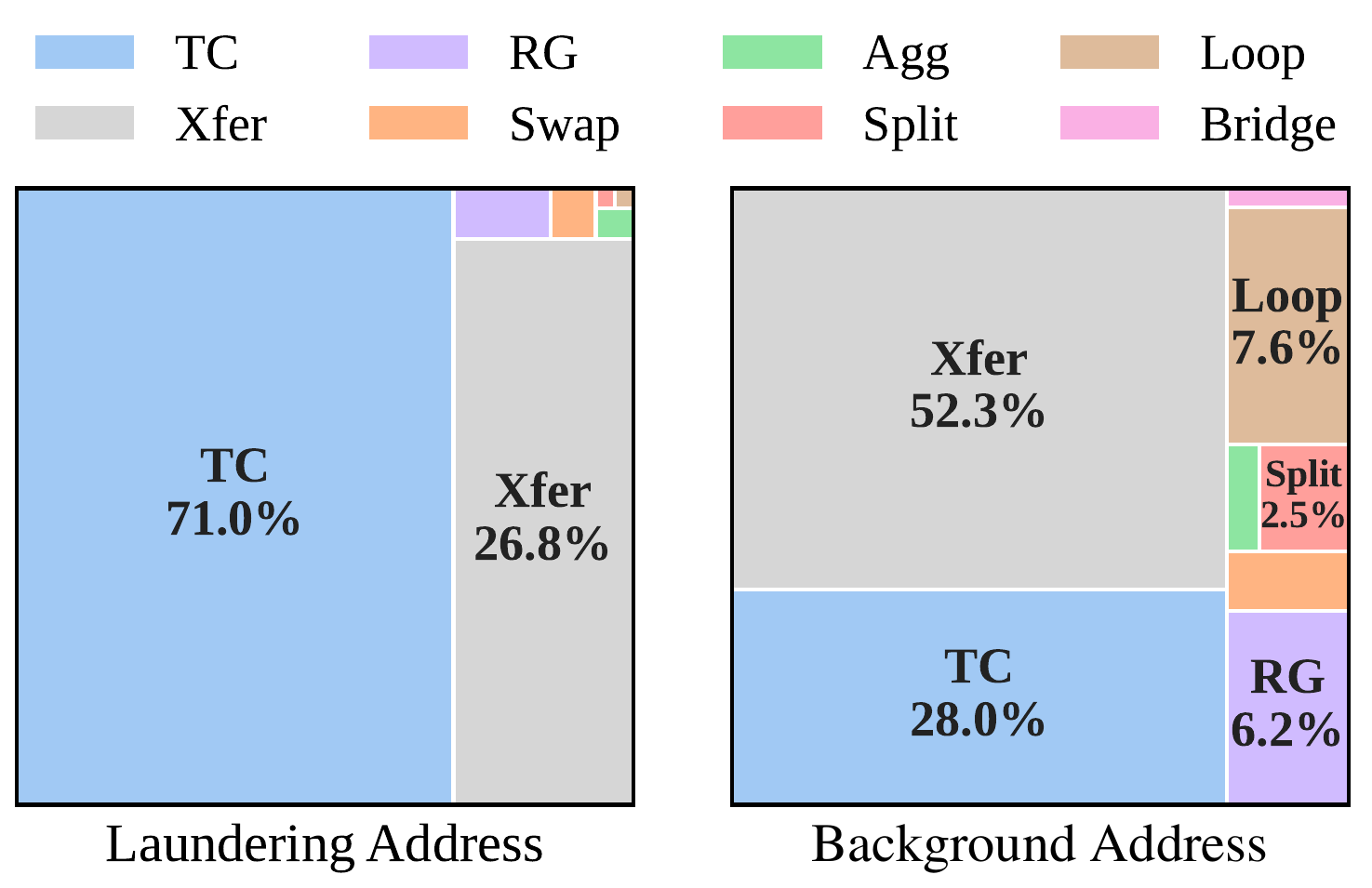}
        \caption{Address-level dominant transaction type.}
        \label{fig:tx-behavior}
    \end{subfigure}
            \hspace{0.03\linewidth}
    \begin{subfigure}[t]{0.44\linewidth}
        \centering
        \includegraphics[width=\linewidth]{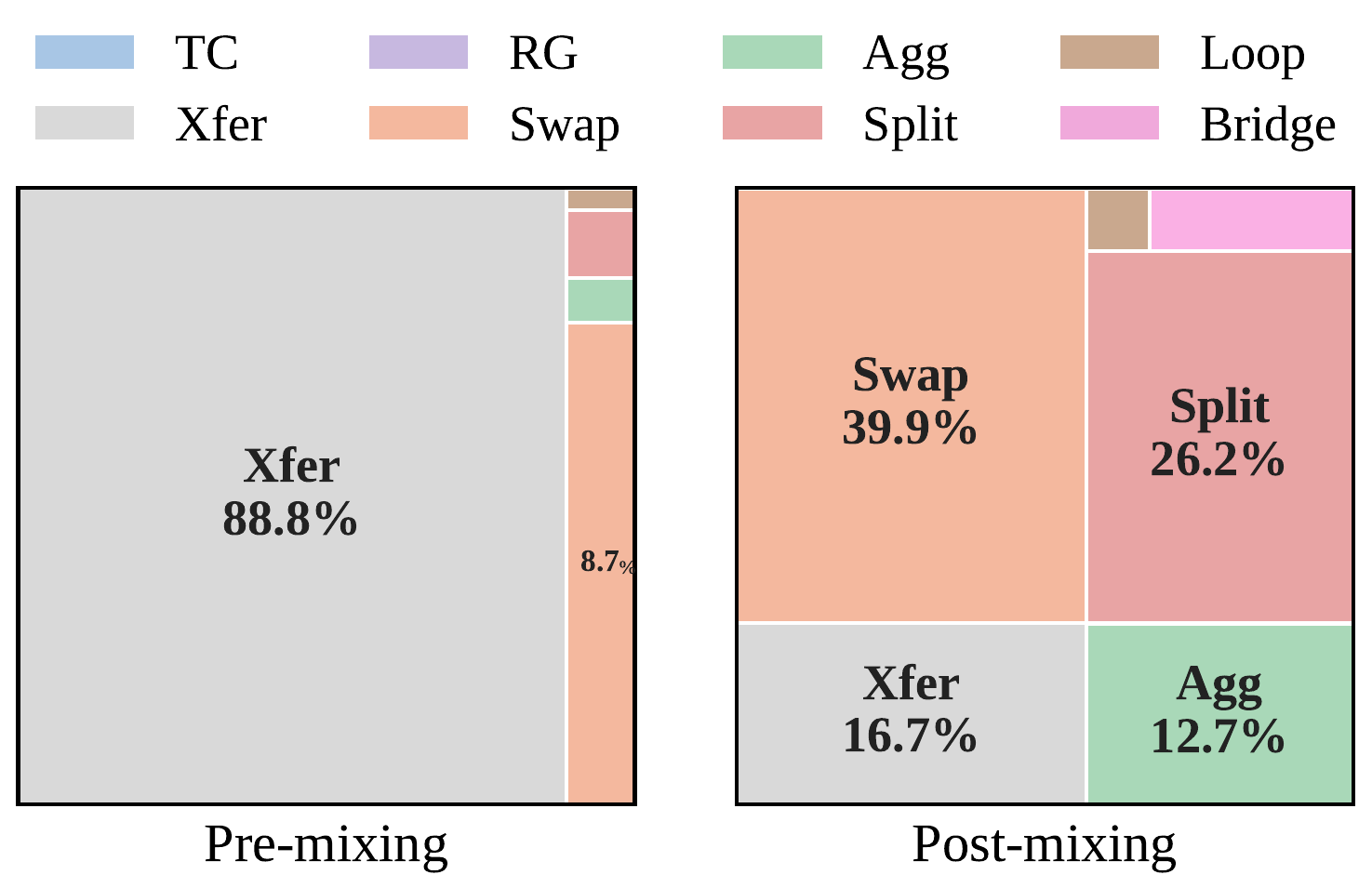}
        \caption{Pre- and post-mixing transaction type.}
        \label{fig:behavior-treemap-up-down}
    \end{subfigure}

    \caption{Dominant transaction types in mixer laundering. 
    (a) Dominant transaction types of laundering and background mixer addresses.
    (b) Transaction types in pre-mixing and post-mixing fund flows.}
    \label{fig:tx-behavior-combined}
\end{figure}

\subsubsection{Role Separation at the Mixer Boundary}
Laundering deposit and withdrawal addresses almost never overlap. Across all cases, 24 of the 27 cases contain no address that serves both deposit and withdrawal roles, and only $1.48\%$ of laundering addresses serve both. This role separation severs the most direct address-level traceability between the deposit and withdrawal. Laundering actors thus avoid the exposure risk of having mixer ingress and egress linked through a shared address.

Laundering deposit and withdrawal addresses also exhibit fundamentally
different behavioral profiles, and both sides deviate substantially from
background mixer addresses. Figure~\ref{fig:address-profile} compares the four
address categories along five dimensions. Laundering deposit addresses display an aggregation structure with high in-degree and low out-degree. They serve as the mixer's fund-aggregation inlets. Laundering withdrawal addresses, in contrast, have little prior on-chain history and short lifespans. $98.6\%$ of laundering withdrawal addresses have at most one day of on-chain activity before their first mixer interaction, indicating that they are disposable one-time outlets.

This functional specialization extends to the dominant transaction types. As shown in Figure~\ref{fig:tx-behavior}, laundering addresses' dominant
transactions are tightly concentrated. \TC interactions and direct transfers together account for $97.8\%$ of dominant types. More complex transaction types are virtually absent. Background mixer addresses show no such concentration. Their dominant behaviors are spread across \TC, direct
transfer, loop, \RG, split, and other types. They more closely resemble
general-purpose wallets used for diverse on-chain activities.

\begin{strategyhead}
\textbf{Strategy 1}: Laundering actors strictly separate deposit and
withdrawal addresses, and the behavioral profiles of both deviate
substantially from those of background mixer addresses.
\end{strategyhead}

\begin{figure}[t]
    \centering
    \begin{subfigure}{0.42\linewidth}
        \centering
        \includegraphics[width=\linewidth]{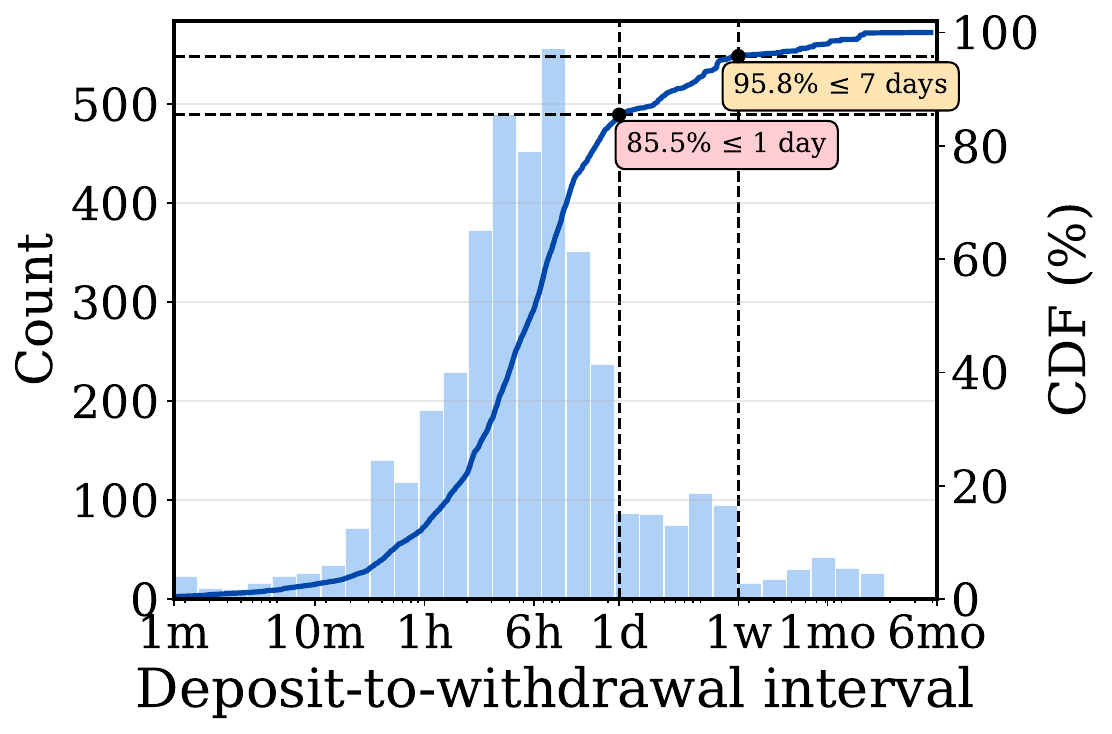}
        \caption{Interval distribution.}
        \label{fig:dw-interval-cdf}
    \end{subfigure}
        \hspace{0.02\linewidth}
    \begin{subfigure}{0.43\linewidth}
        \centering
        \includegraphics[width=\linewidth]{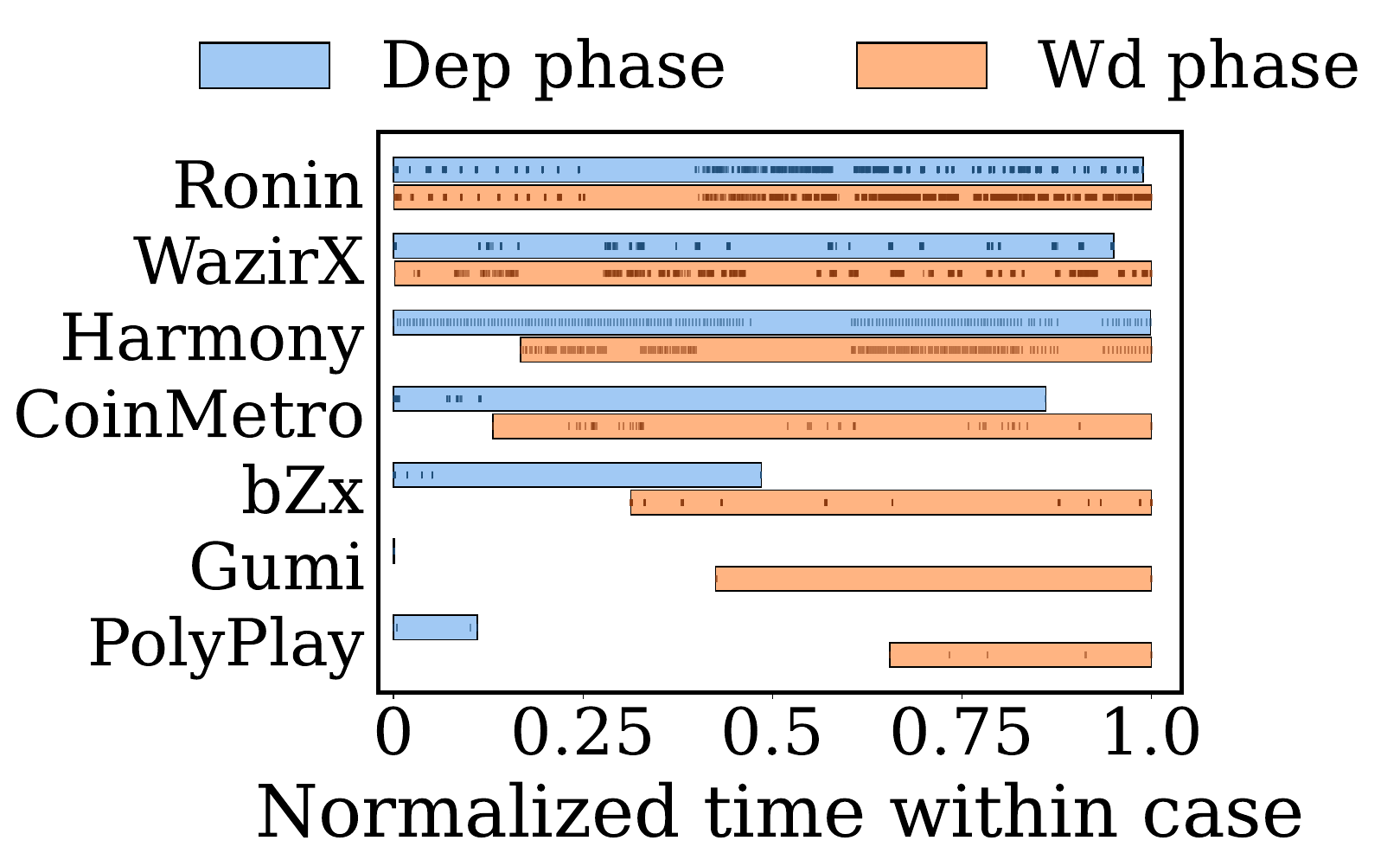}
        \caption{Deposit and withdrawal phase overlap.}
        \label{fig:dw-phase-overlap}
    \end{subfigure}
    \caption{Timing characteristics of laundering deposit and withdrawal.
    (a) Histogram and CDF of the interval between deposit and withdrawal,
    showing rapid turnover. (b) Five representative cases sorted by
    interleaving degree from Ronin to bZx and two strict-sequential cases.}
    \label{fig:timing-overview}
\end{figure}




\subsubsection{Rapid and Interleaved Mixer Use}
In laundering cases, the mixer is used as a rapid transit channel rather than a long-term holding container.  We order deposits and withdrawals within each case and mixer pool by time and measure the intervals between temporally adjacent deposits and withdrawals. 
As shown in Figure~\ref{fig:dw-interval-cdf}, about $85.5\%$ of these intervals are within one day, and $95.8\%$ are within seven days, with the tail beyond one week below $5\%$. The median interval falls within one day in 19 of 27 cases. These short intervals compress the window available for on-chain monitoring, making it difficult to trace the funds before leaving the mixer.

Deposit and withdrawal activity also interleave at the case level. Overall, 25 cases follow the interleaved pattern, with only 2 strictly sequential.
Interleaving destroys the one-to-one temporal correspondence between
deposits and withdrawals. 
Figure~\ref{fig:dw-phase-overlap} shows five representative interleaved
cases and two strict-sequential cases. Among the
five interleaved cases, the top three show deposit and withdrawal phases
that overlap almost entirely. The bottom two only partially overlap. In
Gumi and PolyPlay, all withdrawals begin only after every deposit has
completed. Even if the timestamp of a particular deposit
is known, the corresponding withdrawal cannot be inferred from the
sequence of withdrawal transactions.

\begin{strategyhead}
\textbf{Strategy 2}: Laundering actors use the mixer as a rapid transit channel
and interleave the deposits and withdrawals rather than completing all
deposits before starting any withdrawal.
\end{strategyhead}


\begin{figure}[t]
    \centering
    \begin{subfigure}{0.4\linewidth}
        \centering
        \includegraphics[width=\linewidth]{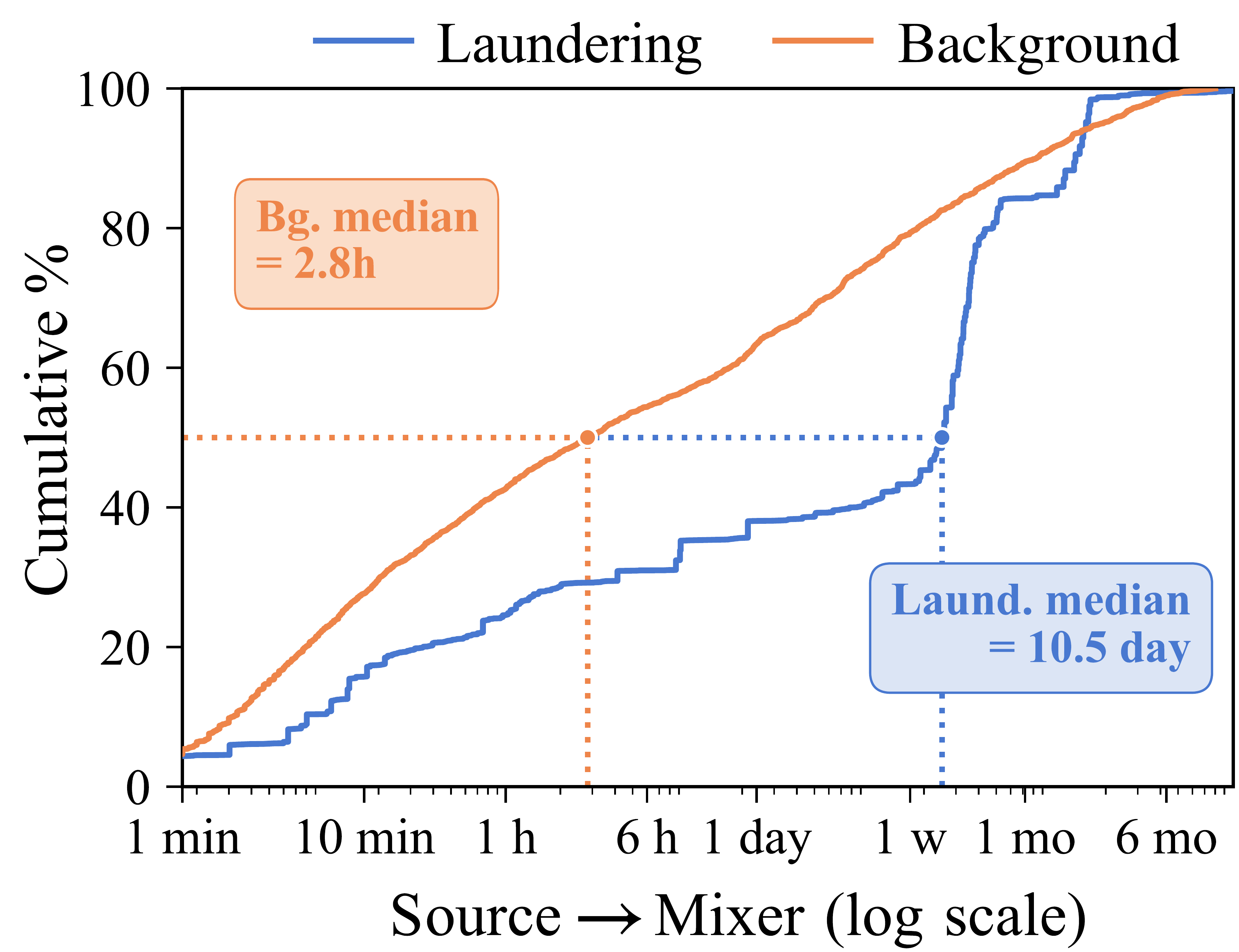}
        \caption{Pre-mixing delay.}
        \label{fig:timing-cdf-upstream}
    \end{subfigure}
        \hspace{0.02\linewidth}
    \begin{subfigure}{0.4\linewidth}
        \centering
        \includegraphics[width=\linewidth]{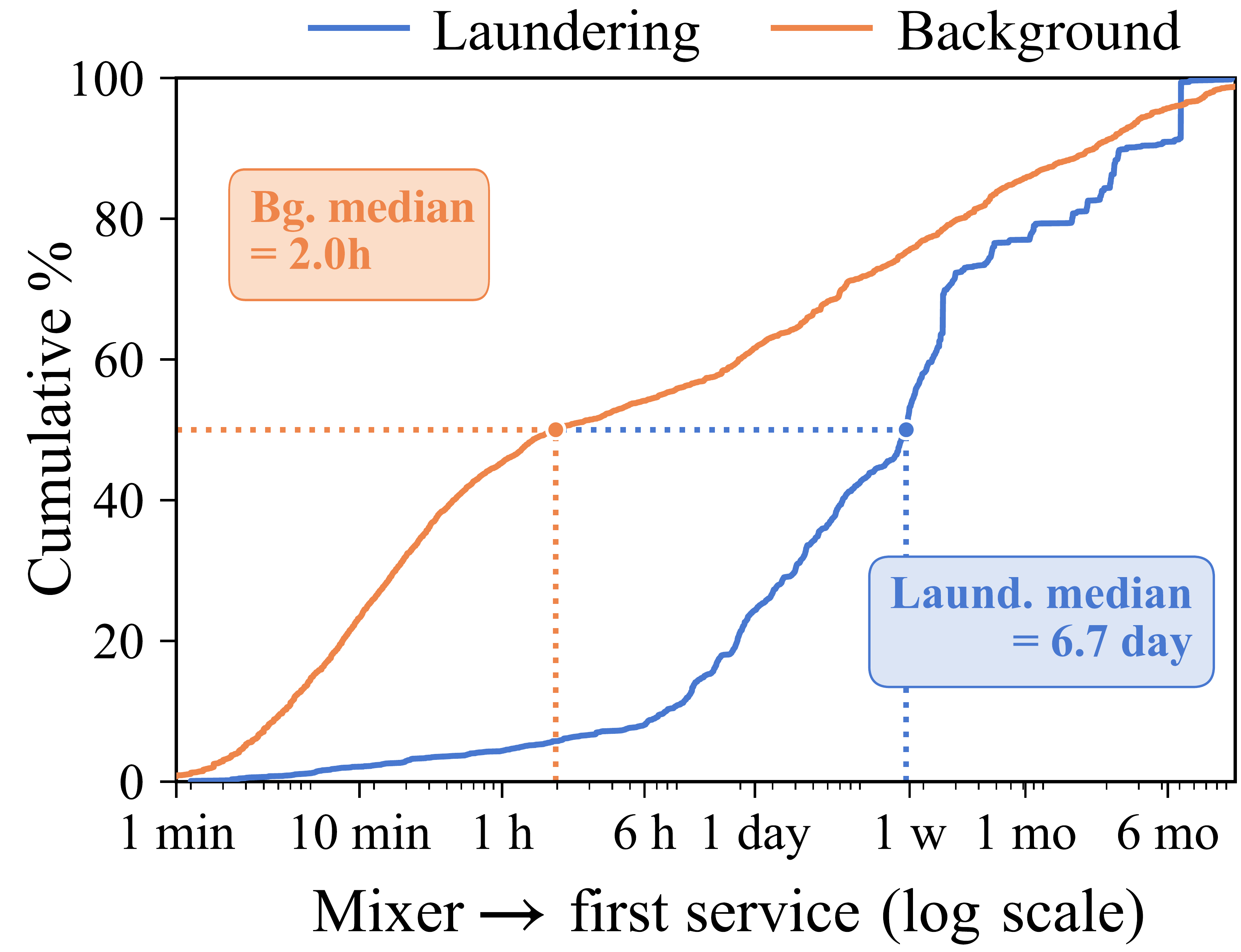}
        \caption{Post-mixing delay.}
        \label{fig:timing-cdf-downstream}
    \end{subfigure}
    \caption{Timing CDFs on a logarithmic scale.
    (a) Time from source transaction to mixer deposit.
    (b) Time from mixer withdrawal to first service destination.}
    \label{fig:timing-cdf}
\end{figure}

\subsubsection{Slow and Concentrated Upstream Preparation}
Illicit funds usually enter the mixer from a single concentrated source. Only $5.2\%$ of laundering deposits are supplied by multiple sources at the immediate upstream hop, compared with $30.3\%$ in background deposits. In 20 of the 27 cases, no more than $10\%$ of deposits are supplied by multiple sources at the immediate upstream hop. 

Illicit funds also move slowly from origin to mixer. As shown in Figure~\ref{fig:timing-cdf-upstream}, the median time from the source to the mixer is $10.5$ days for illicit funds and only $2.8$ hours for background ones. A substantial fraction of illicit funds are not pushed into the mixer immediately. They are held or reorganized for some time before being deposited into the mixer. This long delay makes it difficult to link a deposit transaction back to its original source based on timing alone.

The dominant upstream behavior remains direct transfer ($88.8\%$), as shown in the left part of Figure~\ref{fig:behavior-treemap-up-down}. Swap accounts for $8.7\%$ and is the most prominent secondary signal. When a swap occurs upstream, it is usually close to the mixer, corresponding to a final step that converts heterogeneous assets into ETH to fit the mixer's accepted denomination.

\begin{strategyhead}
\textbf{Strategy 3}: Laundering actors push illicit funds toward the
mixer along a single dominant chain at a slow pace, rather than
aggregating from multiple sources.
\end{strategyhead}





\subsubsection{Post-Withdrawal Fund Restructuring}
Illicit funds do not settle quickly after leaving the mixer. As shown in Figure~\ref{fig:timing-cdf-downstream}, the median time from the mixer withdrawal to the first service destination is $6.7$ days for illicit funds and only $2$ hours for background ones. Most illicit funds therefore undergo a substantial post-withdrawal reorganization period before approaching any identifiable service.


Downstream behavior is dominated by restructuring rather than direct forwarding. As shown in Figure~\ref{fig:behavior-treemap-up-down} (right), swap accounts for $39.9\%$ of laundering downstream behavior, compared with only $1.6\%$ of background activity. Split and aggregation account for $26.2\%$ and $12.7\%$, respectively, while direct transfer accounts for only $16.7\%$. Across cases, these three restructuring behaviors account for a median of $64.2\%$ of downstream activity and collectively exceed direct transfers in 19 of the 27 cases. Illicit funds are thus repeatedly converted, split, and recombined after withdrawal, forming an active restructuring stage rather than being forwarded directly to the next destination.

\begin{strategyhead}
\textbf{Strategy 4}: After leaving the mixer, laundering actors continue
to reshape funds rather than forwarding them directly toward a final
destination.
\end{strategyhead}

\subsubsection{Locally Tight but Globally Loose Case Coordination}
Large-scale mixer laundering involves multiple addresses operating within the same case. These addresses exhibit coordinated behavior through shared upstream counterparties, shared downstream counterparties, and concentrated transaction timing. As shown in Figure~\ref{fig:within-case-shared-cp}, on average,
$90.1\%$ of deposit addresses in a case have at least one deposit whose
immediate upstream sender also funds another deposit address. Similarly,
$68.1\%$ of withdrawal addresses have at least one withdrawal whose
immediate downstream receiver also receives funds from another withdrawal
address. The most active 1-hour window in each case contains, on average,
$33.3\%$ of its deposit transactions. Together, these shared counterparties
and temporal concentration reveal tight coordination within local activity
bursts, even without direct transactions among the involved addresses.

This coordination does not span the full case lifetime. The deposit activity of a case extends over a median of $18.8$ days, with the longest reaching $329$ days, and each case spans an average of $35.7$ distinct 1-hour active windows. Among cases with multiple active windows, the median shares of window pairs sharing no mixer-side address are $68.0\%$ for deposits and $67.6\%$ for withdrawals. Thus, addresses active in one burst often do not reappear in another, weakening address-level continuity across the case lifetime. This makes address-based grouping prone to fragmentation.

\begin{strategyhead}
\textbf{Strategy 5}: Within a case, laundering transactions form tightly
coordinated local bursts, while address-level connections weaken across
bursts over the full case lifetime.
\end{strategyhead}

\subsection{Limitations of Existing Countermeasures}
\label{sub-flaws}

Existing countermeasures against mixer laundering operate at two levels.
Mixer ecosystems may provide compliance mechanisms to screen illicit
activity, while investigators use address-level heuristics to recover
candidate links across the mixer boundary. We evaluate both types of
countermeasures using \textsc{MixLaunder} and identify three limitations.



\subsubsection{Absence of Protocol-Enforced Risk Screening in TC}\TC provides an optional Compliance Tool that allows depositors to generate verifiable reports linking their deposits and withdrawals, but the tool does not assess the risks associated with the funds.
At the pool-contract level, deposits undergo no source-of-funds screening, withdrawals are executed once the cryptographic validity conditions are satisfied, and the protocol does not respond to anomalous activity. For example, in the Harmony case, 4 withdrawal addresses collectively use only 2 relayers across 8 withdrawal transactions. The relayer fee averages about 81 ETH per withdrawal, leaving only 19 ETH for the recipient. Such unusually high fees may indicate collusion between the relayers and laundering addresses, yet \TC raises no risk alert.



\begin{flawhead}
\textbf{Limitation 1}:
\TC's Compliance Tool only supports voluntary verification of deposit--withdrawal links. Its pool contracts neither perform risk screening nor respond to anomalous transactions or relayer behavior.
\end{flawhead}







\subsubsection{Deposit-Only Compliance Screening in RG}\RG adopts compliance via Private Proof of Innocence (POI)~\cite{railgun-ppoi},
which checks each deposit against external blacklists but performs no
check on withdrawals. POI maintains only a flagged-deposit list\footnote{\url{https://ppoi.info/flagged-shields}.}. As a result, the
same address often receives inconsistent POI verdicts across
different transactions. We queried POI status%
\footnote{\url{https://ppoi.info/}.} for the three \RG cases. Among the 37 laundering deposit addresses,
28 ($75.68\%$) have no deposits flagged by POI. Of the remaining nine
addresses, seven have all their deposits flagged, while two receive
mixed verdicts, with earlier deposits unflagged before later ones are
flagged. We further observe two patterns from these results.


\vspace{2pt}
\noindent\textit{{Same address, earlier deposits are unflagged while later ones are banned.}}
$18.92\%$ of deposit addresses have all deposits flagged by POI. For
another $5.41\%$, earlier deposits remain unflagged while later ones are banned.
This reflects two issues: blacklist updates lag, and POI does not
retroactively re-check earlier deposits. 

\vspace{2pt}
\noindent\textit{{Same address, deposits flagged but withdrawals released.}}
Of the nine \RG deposit addresses with at least one flagged
deposit, two (\href{https://etherscan.io/address/0x6c4f805335e6b91350ff291abc1c045971c2696f}{0x6c4f} and \href{https://etherscan.io/address/0xea5f017ceaa2f4fe86a0c1abf1a0c87fb0e5d7fc}{0xea5f})
received another withdrawal before their flagged deposits were
refunded. The prior deposit-side flags therefore do not prevent funds
from leaving the mixer via the same address.

These two patterns show that blacklist-based compliance is
easily bypassed~\cite{dethective2025}: updates lag, are not retroactive, and cover only the deposit side. 
This falls short of the FATF Travel Rule (Recommendation~16)~\cite{fatf-rec16}: the originator and the beneficiary of a value transfer must be assessed under the same risk standard.

\begin{flawhead}
\textbf{Limitation 2}: 
Railgun checks deposits against blacklists but not withdrawals. The same address can have one deposit flagged and another not, or flagged deposits but clean withdrawals, falling short of the FATF Travel Rule's bidirectional coverage.

\end{flawhead}

\subsubsection{Breakdown of Existing Heuristics}
\label{sub-heu}


Address-level deanonymization heuristics exploit observable address reuse and transactional relationships to relink some of the deposit and withdrawal activities concealed by mixers, thereby narrowing the set of candidate deposit--withdrawal pairs. Three commonly used heuristics are: \emph{double role}~\cite{wang2023zkmixer}, where the same address acts as both a deposit and withdrawal address; \emph{direct linkage}~\cite{tc-clustering}, where a deposit address and a withdrawal address transact directly; and \emph{deposit-as-initiator}~\cite{wang2023zkmixer}, where a deposit address initiates a withdrawal transaction. In AML investigations, if an address on one side of a mixer has already been associated with illicit funds, these heuristics can help identify potentially related addresses on the other side and restore the tracing path across the mixer. We therefore examine whether these heuristics can recover deposit--withdrawal linkages in our 27 laundering cases.

Table~\ref{tab:heuristic-breakdown} reports the address-level hit rate and cross-case coverage of each heuristic. Double role, direct linkage, and deposit-as-initiator flag only 1.48\%, 1.27\%, and 0.64\% of mixer-related laundering addresses, respectively, and occur in only 3, 4, and 2 of the 27 cases. These results indicate that most laundering addresses do not exhibit behavioral patterns that expose their deposit--withdrawal relationships, while the few hits are concentrated in a small number of cases. These heuristics capture occasional address reuse, direct interaction, or exposure of the withdrawal initiator, rather than behavioral patterns that consistently occur across cases. Therefore, their limited coverage makes them insufficient as standalone methods for tracing laundering funds across mixers.


\begin{flawhead}
\textbf{Limitation 3}: Three deanonymization heuristics
that link mixer deposits to withdrawals flag a few laundering addresses, because they rely on operational mistakes that disciplined laundering does not make.
\end{flawhead}

\begin{table}[t]
\centering
\caption{Address hit rates and cross-case coverage of three representative address-level heuristics.}
\label{tab:heuristic-breakdown}
\small
\resizebox{0.9\textwidth}{!}{%
\begin{tabular}{lccl}
\toprule
\textbf{Heuristic} & \textbf{Addr. Hit Rate} & \textbf{Case Coverage} & \textbf{Observed Limitation} \\
\midrule
Double role          & $1.48\%$ & $3/27$ & Most deposit and withdrawal addresses are distinct. \\
Direct linkage       & $1.27\%$ & $4/27$ & Direct transactions between the two sides are rare. \\
Deposit-as-initiator & $0.64\%$ & $2/27$ & Deposit addresses rarely initiate withdrawals. \\
\bottomrule
\end{tabular}
}
\end{table}

\section{\textsc{MixGuard:}~Detection Framework}
\label{sec-framework}

We propose \textsc{MixGuard}, which combines a tri-view laundering encoder with a two-stage grouping pipeline to identify laundering-related mixer transactions and organize them into same-case groups. We first explain the design requirements and how our measurement findings guide the framework~(\S\ref{subsec:motivation}). We then describe how the encoder learns transaction representations from address, path, and subgraph evidence~(\S\ref{subsec-trile}), how the strong-tie clusterer forms reliable local cores~(\S\ref{subsec-STC}), and how the weak-tie merger links related cores to improve same-case recovery~(\S\ref{subsec-wtm}). The feature groups used by \textsc{MixGuard} are summarized in Table~\ref{tab:mixguard-features} in
Appendix~\ref{app:exp-config}.

\subsection{Design Motivation}
\label{subsec:motivation}

Our measurement analysis in \S\ref{sec-mixerFeature} shows that a practical mixer-laundering detection framework should satisfy two requirements: distinguishing laundering transactions from background mixer usage without relying on exact deposit--withdrawal links, and organizing suspicious transactions distributed across different addresses, stages, and time windows into same-case groups. These requirements lead to three design implications.

{First}, Strategies~1--4 show that mixer-laundering evidence spans address behavior, timing, fund-flow paths, and local topology, making any single view insufficient. We therefore design a \textbf{\emph{tri-view laundering encoder}} that integrates address, path, and topology information. Joint detection and case-aware objectives enable the learned representation to distinguish laundering from background mixer activity while preserving same-case relationships.
{Second}, Strategy~5 shows that same-case transactions exhibit strong relationships within local activity bursts, such as close execution times and shared upstream or downstream counterparties. We therefore use a \textbf{\emph{strong-tie clusterer}} to organize highly related suspicious transactions into small, reliable core clusters, reducing background contamination and cross-case merging.
{Third}, Strategy~5 also shows that these local bursts retain only weak relationships over the full case lifetime because they may involve different addresses and occur across different laundering stages and time windows. Relying only on strong ties therefore leaves same-case activity fragmented. We use a \textbf{\emph{weak-tie merger}} to learn weak associations between core clusters, improving same-case recovery coverage while limiting cross-case merging.

\subsection{Tri-view Laundering Encoder}
\label{subsec-trile}

To jointly represent the laundering signals scattered across multiple views, we design the \emph{Tri-view laundering encoder}. It encodes evidence from the address, path, and subgraph views into a shared representation. Two task heads, a detection head and a projection head, are then jointly trained on this representation.

\underline{\emph{Address view}} characterizes the behavioral profile of each deposit or withdrawal address, capturing address-level differences in counterparty distribution, temporal behavior, and service-interaction patterns. The address feature $\mathbf{x}_a$ is encoded by an MLP~\cite{tolstikhin2021mlp} into the address embedding:
\begin{equation}
\mathbf{h}_a = \mathrm{MLP}_{\mathrm{addr}}(\mathbf{x}_a) \in \mathbb{R}^{d}.
\label{eq:addr}
\end{equation}

\underline{\emph{Path view}} summarizes the upstream and downstream fund-flow behavior of a mixer transaction, capturing path length, cross-service routing, and more. This view complements the address-level profile by capturing the fund preparation behavior for deposits and the fund reshaping behavior for withdrawals. The path feature $\mathbf{x}_p$ is encoded by an  MLP into the path embedding:
\begin{equation}
\mathbf{h}_p = \mathrm{MLP}_{\mathrm{path}}(\mathbf{x}_p) \in \mathbb{R}^{d}.
\label{eq:path}
\end{equation}


\underline{\emph{Topology view}} focuses on the structure of fund flow, targeting structural signals such as aggregation and dispersion.
Two-layer GraphSAGE~\cite{hamilton2017inductive} is used to propagate and aggregate neighborhood features, yielding node representations
\begin{equation}
\mathbf{H}^{(\ell+1)} = \sigma\!\left(\mathbf{W}^{(\ell)} \left[\mathbf{H}^{(\ell)} \,\|\, \mathrm{AGG}_{\mathcal{N}}(\mathbf{H}^{(\ell)})\right]\right)
\end{equation}
for $\ell \in \{0, 1\}$, where $\mathrm{AGG}_{\mathcal{N}}$ is the mean aggregation function, $\mathbf{H}^{(0)}$ contains the address structural features, and $\mathbf{H}^{(2)}$ is the final output.

To prevent the mixer transactions' own signal from being diluted by its neighbors, we concatenate the mean-pooled subgraph representation with the embedding $\mathbf{h}_v^{(2)}$ and project, yielding the topology
embedding:
\begin{equation}
  \mathbf{h}_t = \mathbf{W}_t [\,\text{Mean-Pool}(\mathbf{H}^{(2)}) \,\|\, \mathbf{h}_v^{(2)}\,] \in \mathbb{R}^{d}.
  \label{eq:topo}
\end{equation}

Because the three views provide complementary evidence rather than redundant views of the same signal, we fuse them via concatenation followed by projection:
\begin{equation}
\mathbf{h}_f = \mathrm{MLP}_{\text{fuse}}([\mathbf{h}_a \,\|\, \mathbf{h}_p \,\|\, \mathbf{h}_t]) \in \mathbb{R}^{d}.
  \label{eq:fused}
\end{equation}
Concatenation lets the projection layer learn cross-view weights without forcing the views onto a common scale. The resulting $\mathbf{h}_f$ serves as the laundering-aware
representation shared by the following two task heads.



\paragraph{Multi-task training.}
To ensure that the learned representation can both separate laundering transactions from background ones and pull together laundering ones from the same case, we jointly train two task heads with the total loss $\mathcal{L} = \mathcal{L}_{\text{BCE}} + \alpha\, \mathcal{L}_{\text{MSupCon}}.$
The detection head applies a standard Binary Cross-Entropy loss~\cite{bishop2006pattern} to separate \modified{labeled laundering transactions from background transactions} in the representation space.
The projection head maps $\mathbf{h}_f$ to an $\ell_2$-normalized embedding
\begin{equation}
  \mathbf{z} = \frac{\mathrm{MLP}_{\text{proj}}(\mathbf{h}_f)}
                    {\|\mathrm{MLP}_{\text{proj}}(\mathbf{h}_f)\|_2}
              \in \mathbb{R}^{d_z},
  \label{eq:proj}
\end{equation}
on which we apply a masked supervised contrastive loss. 
Unlike the standard
supervised contrastive loss~\cite{khosla2020supervised}, only same-case laundering transactions are
treated as positives, while laundering transactions from other cases and all
background transactions serve as negatives:
\begin{equation}
  \mathcal{L}_{\text{MSupCon}}
  = -\frac{1}{|\mathcal{B}_L|}
    \sum_{i \in \mathcal{B}_L}
    \frac{1}{|P(i)|}
    \sum_{p \in P(i)}
    \log
    \frac{\exp(\mathbf{z}_i^\top \mathbf{z}_p / \tau)}
         {\sum_{a \in A(i)} \exp(\mathbf{z}_i^\top \mathbf{z}_a / \tau)},
  \label{eq:msupcon}
\end{equation}
where $\mathcal{B}$ is the training batch, $\mathcal{B}_L \subseteq \mathcal{B}$
its laundering subset, $c(i)$ the case label of transaction $i$, $\mathbf{z}_i$
the projection-head output of transaction $i$ defined by Eq.~(\ref{eq:proj}),
and $\tau$ a temperature hyperparameter. The positive set
$P(i) = \{p \in \mathcal{B}_L \mid c(p) = c(i),\, p \neq i\}$ collects same-case
laundering transactions, while the candidate set
$A(i) = \mathcal{B} \setminus \{i\}$ covers all other transactions in the batch.


\subsection{Strong-Tie Clusterer}
\label{subsec-STC}

Mixer laundering transactions within the same case are not necessarily pairwise highly similar: across different time windows and laundering tactics, only weak signals tend to remain between them, while direct global clustering is highly sensitive to noise and to coincidental cross-case similarity. We therefore adopt a two-stage strategy: \emph{Strong-tie clusterer} first discovers tightly bonded core subgroups by retaining only high-confidence strong-similarity relations; \emph{Weak-tie merger} then learns weak-association evidence at the cluster-pair level to further merge subgroups that belong to the same case but are connected only by weak signals.



Classical clustering methods are ill-suited to \emph{strong-tie clusterer}: the number of laundering cases is unknown and keeps growing, and laundering-group structures vary widely across cases, ruling out K-means~\cite{lloyd1982least} with its preset cluster count, GMM~\cite{reynolds2009gaussian} with its shape assumption, and DBSCAN~\cite{ester1996density} with its global density threshold. This {module} therefore aims to discover tightly bonded, strong-similarity subgroups directly from the projection embeddings $\mathbf{z}$ produced by \textit{tri-view encoder}, without presetting the cluster number or assuming any data distribution. 

Since $\mathbf{z}$ is already $\ell_2$-normalized, we measure the cosine similarity between two transactions and select its top-$k$ nearest neighbors: $\mathrm{sim}(i, j) = \mathbf{z}_i^{\top}\mathbf{z}_j \in [-1, 1].$
For each transaction $i$ we take its top-$k$ nearest neighbors
\begin{equation}
  \mathcal{N}_k(i) =
  \arg\!\max\nolimits_{\,j \in \mathcal{S} \setminus \{i\}}^{(k)}\,
  \mathrm{sim}(i, j),
  \label{eq:topk}
\end{equation}
and construct the strong-tie edge set by retaining the pairs that are mutual nearest neighbors and whose similarity is at least $\tau_s$:
\begin{equation}
  \mathcal{E}_{\mathrm{strong}} =
  \big\{\,(i,j)\,\big|\,
        j \in \mathcal{N}_k(i)\,\wedge\,
        i \in \mathcal{N}_k(j)\,\wedge\,
        \mathrm{sim}(i,j) \geq \tau_s\,\big\}.
  \label{eq:edge}
\end{equation}
The mutual-neighbor condition rules out asymmetric edges where $j$ is among the top-$k$ neighbors of $i$ but $i$ is not among those of $j$. 
Background transactions are dispersed across the projection space by the masked supervised contrastive loss. They rarely form stable mutual neighbors with laundering transactions or exceed $\tau_s$, and are therefore not added to the edge set.

Finally, we construct an undirected graph $\mathcal{G}_{\mathrm{strong}}$ with mixer transactions as nodes and $\mathcal{E}_{\mathrm{strong}}$ as edges, and take its connected components $\{C_1, \ldots, C_M\} = \mathrm{ConnComp}\big(\mathcal{G}_{\mathrm{strong}}\big)$,
each $C_m$ being a \emph{strong-tie cluster} that serves as the merging unit for the \emph{Weak-tie merger} in Sec.~\ref{subsec-wtm}.

\subsection{Weak-Tie Merger}
\label{subsec-wtm}

After \emph{strong-tie clusterer}, transactions from the same laundering case may still be split across several strong-tie clusters. 
This is because a single case often spans a long time window and combines multiple laundering tactics, so its subgroups end up linked only by weak signals that the strong-similarity edges of \emph{strong-tie clusterer} cannot recover. We therefore introduce \emph{Weak-tie merger}, which learns a cluster-pair classifier for whether two strong-tie clusters belong to the same case, and merges those connected only by weak signals into \emph{case-level} clusters.

For any two strong-tie clusters $(C_m, C_n)$ produced by \textit{strong-tie clusterer}, we extract a pair-level feature vector $\mathbf{f}_{m,n}$ that summarizes cross-cluster similarity, structural and service consistency, and scale and temporal alignment. We then label two strong-tie clusters from the same ground-truth case as a positive pair and any cross-case combination as a negative pair, and train a LightGBM\footnote{It trains stably on features of heterogeneous scales without requiring normalization.} binary classifier~\cite{ke2017lightgbm} to predict whether each pair should be merged. We then apply union-find on the predicted-merge pairs, yielding a set of case-level clusters $\{\widetilde{C}_1, \ldots, \widetilde{C}_K\}$ with $K \leq M$.

\section{Evaluation of \textsc{MixGuard}}
\label{sec-evaluation}










To comprehensively evaluate \textsc{MixGuard}, we first describe the experimental setup~(\S\ref{subsec:setup}). We then compare its transaction-level detection and case-aware grouping performance with representative baselines~(\S\ref{sub-per}), analyze the contribution of each component~(\S\ref{subsec:ablation}), present case studies of the laundering stages captured by recovered groups~(\S\ref{subsec:group-dis}), and assess its ability to discover previously unseen incidents in the wild~(\S\ref{subsec:wild}).

\subsection{Experimental Setup}
\label{subsec:setup}

\begin{table}[t]
\centering
\setlength{\tabcolsep}{6pt}
\caption{Comparison results between \textsc{MixGuard} and baselines on Transaction~(Tx)-level Detection and Case-aware Grouping performance on nine cross-validation folds.}
\label{tab:main_results_four_metrics}
\resizebox{0.8\textwidth}{!}{%
\begin{tabular}{llcccc}
\toprule
\multirow{2}{*}{Type} & \multirow{2}{*}{Method}
& \multicolumn{2}{c}{Tx-level Detection}
& \multicolumn{2}{c}{Case-aware Grouping} \\
\cmidrule(lr){3-4} \cmidrule(lr){5-6}
& & AUROC(\%) $\uparrow$ & Prec(\%) $\uparrow$
& Purity(\%) $\uparrow$ & Top10-Cov(\%) $\uparrow$ \\
\midrule

\multirow{2}{*}{CO}
& K-means
& 68.77$_{\textcolor{blue}{\blacktriangledown 26.69\%}}$
& 19.92$_{\textcolor{blue}{\blacktriangledown 79.20\%}}$
& 51.72$_{\textcolor{blue}{\blacktriangledown 47.44\%}}$
& -- \\
& HDBSCAN
& \underline{92.48}$_{\textcolor{blue}{\blacktriangledown 1.42\%}}$
& 95.30$_{\textcolor{blue}{\blacktriangledown 0.48\%}}$
& 91.06$_{\textcolor{blue}{\blacktriangledown 7.46\%}}$
& 71.67$_{\textcolor{blue}{\blacktriangledown 25.30\%}}$ \\
\midrule

\multirow{2}{*}{DTC}
& MLP
& 86.43$_{\textcolor{blue}{\blacktriangledown 7.87\%}}$
& 85.38$_{\textcolor{blue}{\blacktriangledown 10.84\%}}$
& 94.52$_{\textcolor{blue}{\blacktriangledown 3.94\%}}$
& 36.30$_{\textcolor{blue}{\blacktriangledown 62.17\%}}$ \\
& LightGBM
& 91.53$_{\textcolor{blue}{\blacktriangledown 2.43\%}}$
& \underline{95.70}$_{\textcolor{blue}{\blacktriangledown 0.06\%}}$
& 93.75$_{\textcolor{blue}{\blacktriangledown 4.73\%}}$
& 27.82$_{\textcolor{blue}{\blacktriangledown 71.01\%}}$ \\
\midrule

\multirow{2}{*}{AML}
& DenseFlow
& 81.87$_{\textcolor{blue}{\blacktriangledown 12.73\%}}$
& 89.38$_{\textcolor{blue}{\blacktriangledown 6.66\%}}$
& 78.95$_{\textcolor{blue}{\blacktriangledown 19.77\%}}$
& 62.41$_{\textcolor{blue}{\blacktriangledown 34.96\%}}$ \\
& MG-HRL
& 61.53$_{\textcolor{blue}{\blacktriangledown 34.41\%}}$
& 29.50$_{\textcolor{blue}{\blacktriangledown 69.19\%}}$
& \underline{97.34}$_{\textcolor{blue}{\blacktriangledown 1.08\%}}$
& 36.49$_{\textcolor{blue}{\blacktriangledown 61.97\%}}$ \\
\midrule

\multirow{2}{*}{MD}
& MixBroker
& 73.58$_{\textcolor{blue}{\blacktriangledown 21.56\%}}$
& 51.42$_{\textcolor{blue}{\blacktriangledown 46.30\%}}$
& 97.09$_{\textcolor{blue}{\blacktriangledown 1.33\%}}$
& \underline{83.39}$_{\textcolor{blue}{\blacktriangledown 13.09\%}}$ \\
& MixLinker
& 68.93$_{\textcolor{blue}{\blacktriangledown 26.52\%}}$
& 36.76$_{\textcolor{blue}{\blacktriangledown 61.61\%}}$
& 97.11$_{\textcolor{blue}{\blacktriangledown 1.31\%}}$
& 82.44$_{\textcolor{blue}{\blacktriangledown 14.08\%}}$ \\
\midrule

\textbf{Ours}
& \textsc{MixGuard}
& \textbf{{93.81}}$_{\textcolor{red}{\blacktriangle 1.44\%}}$
& \textbf{{95.76}}$_{\textcolor{red}{\blacktriangle 0.06\%}}$
& \textbf{{98.40}}$_{\textcolor{red}{\blacktriangle 1.09\%}}$
& \textbf{{95.95}}$_{\textcolor{red}{\blacktriangle 15.06\%}}$ \\
\bottomrule
\end{tabular}
}
\par
{\footnotesize\raggedright
Bold denotes the best results, and underlined denotes the
second-best results.\par
``--'' indicates K-means preset the cluster number~$(k{=}3)$,
and is therefore excluded from the Top10-Cov. computation.\par
$\textcolor{red}{\blacktriangle\%}$ on the best result indicates
its relative improvement over the second-best,
$\textcolor{blue}{\blacktriangledown\%}$ on every other cell
indicates its relative gap to the best.\par}
\end{table}

\noindent\textbf{Baselines.}
We select baselines along two dimensions, methodology and domain. \emph{On methodology}, we compare against generic clustering and against the detection-then-clustering pipeline, to assess the necessity of jointly training detection and case-aware grouping in the same representation. \emph{On domain}, we compare against existing mixer deanonymization and general Anti-Money Laundering methods, to verify that \textsc{MixGuard}'s mixer-laundering-specific design is necessary:

\begin{packeditemize}
\item \textit{Cluster-only~(CO)}: classical clustering methods, including
      the K-means~\cite{lloyd1982least} and HDBSCAN~\cite{campello2013density}, applied directly to the concatenated address and path features. 

\item \textit{Detect-then-cluster~(DTC)}: two representative supervised classifiers,
       MLP~\cite{tolstikhin2021mlp} and LightGBM~\cite{ke2017lightgbm}, trained on address and path features for laundering detection, each followed by HDBSCAN for case-level grouping. 
\item \textit{Anti-money laundering~(AML)}: DenseFlow~\cite{lin2024denseflow},
      a dense-subgraph mining approach;
      MG-HRL~\cite{li2025multi}, a multi-view graph-based hierarchical learning method.
      \item \textit{Mixer deanonymity~(MD)}: MixBroker~\cite{du2024breaking},
      a graph-feature-learning framework for breaking the anonymity of
      mixers; MixLinker~\cite{Wang2025improving}, a Personalized
      PageRank-based address linking method.
\end{packeditemize}


\vspace{2pt}
\noindent\textbf{Evaluation Setting.}
To evaluate generalization to unseen laundering cases, we require the
cases in the train, validation, and test sets to be completely
disjoint. With 27 laundering cases in the dataset~\textsc{MixLaunder}, we adopt a 9-fold case-level holdout:
each fold partitions the cases into roughly 20 training, 4 validation, and 3 test cases, and this design ensures every case is tested exactly once, so all cases participate in evaluation. 
As background data, we collect background mixer transactions from TC and
RG within the same time windows as the laundering cases, and
include them alongside the laundering data in the train,
validation, and test splits. All metrics are computed independently per fold and averaged.
\modified{All baselines are evaluated using the same nine
case-level splits and the same transaction features available to each method. We adapt their native outputs to the two evaluation tasks: transaction scores are converted into binary alerts using a threshold of $0.5$, while predicted links or subgraphs are converted into transaction groups for case-aware evaluation. Detailed configurations are provided in Appendix~\ref{app:exp-config}.}

\vspace{2pt}
\noindent\textbf{Evaluation metrics.}
We evaluate \textsc{MixGuard} at two levels: whether it can identify laundering-related mixer transactions, and whether it can organize them into case-level groups.
\begin{packeditemize}
\item \textit{Tx-level detection} evaluates the reliability of suspicious-transaction alerts. \texttt{AUROC} reflects the overall separability between laundering and background transactions, and {Precision}~(\texttt{Prec}) measures the fraction of true laundering transactions among all predicted laundering ones. The higher \texttt{Prec}, the fewer false alarms and the less analyst review effort. \modified{We use a fixed decision threshold of $0.5$ to convert the predicted laundering probability into a binary label.}
\item \textit{Case-aware grouping} evaluates the quality of recovered laundering cases.
\texttt{Purity} measures whether each predicted group mainly contains transactions from the same case, while Top-10 Cluster Coverage~(\texttt{Top-10-Cov}) measures how much of each case can be covered by its top-10 predicted groups.
\end{packeditemize}



\vspace{2pt}
\noindent\textbf{Experimental configurations.}
We implement \textsc{MixGuard} in PyTorch and PyTorch Geometric and evaluate all methods under the same 9-fold case-level holdout splits. Unless otherwise stated, \textsc{MixGuard} uses a 128-dimensional tri-view representation, loss
weight $\alpha{=}0.5$, and the clustering parameters $k{=}20$ and $\tau_s{=}0.90$. Hyperparameter sensitivity analyses are
reported in Appendix~\ref{subsec:hyper} and full configuration details are reported in Appendix~\ref{app:exp-config}.





\subsection{Performance Comparisons}
\label{sub-per}
We evaluate \textsc{MixGuard} against eight baselines on \textsc{MixLaunder} dataset across nine cross-validation folds.
\modified{Table~\ref{tab:main_results_four_metrics} compares \textsc{MixGuard} with eight baselines on transaction-level detection and case-aware grouping.
Appendix~\ref{subsec:A-results} provides detailed nine-fold results and standard deviations in
Table~\ref{tab:mixguard-nine-fold}, and performance by case size in Table~\ref{tab:mixguard-case-size}.
Based on these results, we make the following observations.}

\vspace{2pt}
\noindent\ding{182}~\textbf{\textsc{MixGuard} achieves the best performance in both detection and grouping.}
\textsc{MixGuard} achieves the balanced and highest results on all four metrics, indicating that the advantage of \textsc{MixGuard} comes from its ability to provide both reliable suspicious-transaction detection and high-quality case-level recovery. Specifically, the high \texttt{AUROC} and \texttt{Prec} show that \textsc{MixGuard} can reliably distinguish laundering transactions from a large background of mixer transactions while reducing false alarms, thereby lowering the manual review burden for analysts. \modified{We further examined background transactions flagged as laundering using public reports and on-chain records, but the available public information was insufficient to determine whether these transactions involved money laundering.}
Meanwhile, the high \texttt{Purity} and \texttt{Top10-Cov} show that the detected suspicious transactions can be organized into clean and well-covered laundering groups. In particular, \textsc{MixGuard} improves \texttt{Top10-Cov} by 15.06\% over the second-best result, demonstrating its advantage in recovering complete laundering cases. \textsc{MixGuard} achieves this by first using strong ties to build reliable cores and then using weak-link expansion to merge them into complete laundering groups. 

\vspace{2pt}
\noindent\ding{183}~\textbf{Existing baselines fail to jointly support detection and case recovery.}
\textbf{CO} methods cluster transactions by similarity, without laundering-aware supervision. K-means therefore fails to separate laundering from background mixer usage, yielding \texttt{Prec} below 20\%, while HDBSCAN's low \texttt{Top10-Cov} shows that it recovers only case fragments.
\textbf{DTC} methods can detect some laundering transactions, as shown by the second-best precision of LightGBM, but its \texttt{Top10-Cov} is below 30\%, indicating that independently classified alerts remain fragmented without same-case relation modeling.
\textbf{AML} methods lack mixer-specific behavior modeling, which limits their performance. MG-HRL achieves high \texttt{Purity} but low \texttt{Prec} and \texttt{Top10-Cov}, suggesting noisy alerts and limited case coverage.
\textbf{MD} methods also suffer from low precision, showing that deposit-withdrawal linkage is unreliable for laundering detection when attackers delay withdrawals and rotate addresses.

\subsection{Component Effectiveness Analysis}
\label{subsec:ablation}



\begin{figure}[t]
  \centering
    \includegraphics[width=0.88\columnwidth]{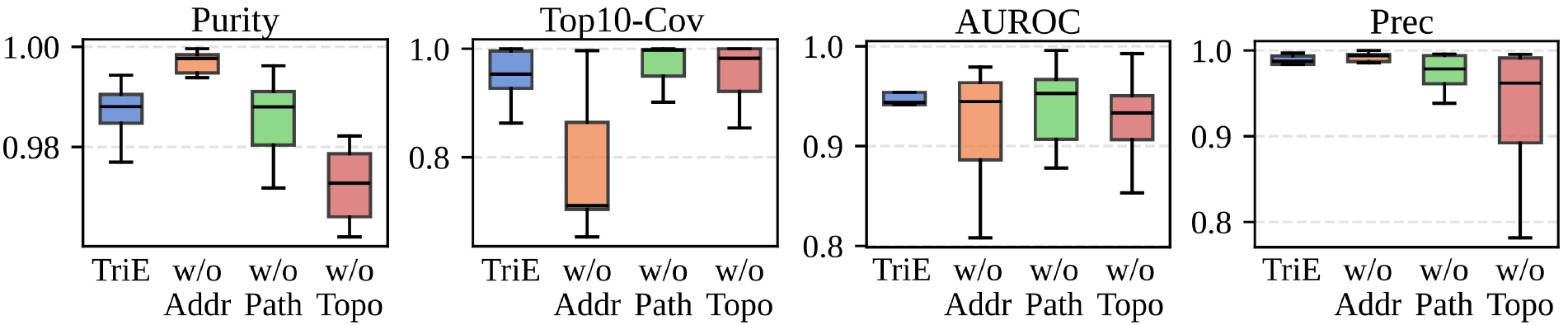}
  \caption{Tri-view ablation across 9 folds. Removing any single view degrades at least one metric.}
  \label{fig:ablation-view}
\end{figure}

\noindent\textbf{Effectiveness of tri-view encoding.}
Figure~\ref{fig:ablation-view} shows that the address, path, and topology views
are complementary. 
Removing the address view mainly
reduces \texttt{Top10-Cov}, indicating that address behavior is important for consolidating same-case transactions. 
Removing the path view increases fold-to-fold variation, showing that upstream/downstream fund-flow context improves robustness across heterogeneous cases.
Removing the topology view lowers \texttt{Prec} and \texttt{Purity}, suggesting that structure helps avoid false positives and cross-case contamination.

\begin{table}[t]
  \centering
  \caption{Real-world mixer laundering incidents in January 2026
  detected by \textsc{MixGuard}, cross-validated against public reports.
  Cause: SC -- Smart-Contract Exploit, HWC -- Hot-Wallet Compromise,
  MSC -- Multisig Compromise, PSB -- Payment-Service Breach.}
  \label{tab:wild_cases}
  \small
  \begin{tabular}{lllccc}
    \toprule
    \textbf{Incident} & \textbf{Cause} & \textbf{Upstream}
    & \# Mixing ETH & \modified{Mixing Date}
    & \modified{Report Date} \\
    \midrule
    Aave~\cite{theblock_multisig}
      & MSC
      & Single direct
      & 1,000
      & Jan 5
      & {Jan 6} \\
    Upbit~\cite{mexc_upbit}
      & HWC
      & Fan in
      & 1,400
      & Jan 6
      & {Jan 8} \\
    Saga~\cite{cryptopolitan_saga}
      & SC
      & Co fund
      & 2,000
      & Jan 24
      & {Jan 25} \\
    Coinbase~\cite{cryptopolitan_coinbase}
      & PSB
      & DEX swap
      & 2,000
      & Jan 26
      & {Jan 27} \\
    \bottomrule
  \end{tabular}
\end{table}


\vspace{2pt}
\noindent\textbf{Effectiveness of strong-tie clusterer.}
Removing the strong-tie filter and directly applying the weak-tie merger to
unfiltered clusters reduces \texttt{AUROC} from 0.938 to 0.837. This shows that
the strong-tie clusterer acts as a precision gate: it forms high-confidence
local cores before weak merging, preventing background transactions from being
mixed into laundering groups.


\vspace{2pt}
\noindent\textbf{Effectiveness of weak-tie merging.}
Removing the weak-tie merger and using only strong-tie clusters reduces
\texttt{Top10-Cov} from 0.960 to 0.868, while increasing the number of clusters
by more than fourfold. This confirms that strong ties alone recover reliable but
fragmented cores and the weak-tie merger is needed to connect these fragments into
case-level laundering groups.

\subsection{Case Studies of Group Recovery}
\label{subsec:group-dis}

Transaction-level detection can identify suspicious mixer transactions, but does not reveal how these transactions form a laundering stage. Group recovery links related transactions, revealing how laundering actors coordinate fund movements into or out of mixers across addresses and time. Following the aggregate evaluation across all 27 cases under case-level holdout, we present three complementary case studies to illustrate the patterns revealed by the recovered groups.

\noindent\textbf{Fantom Foundation: parallel, multi-denomination mixer entry.}
In the Fantom Foundation case, \textsc{MixGuard} groups 140 deposits made by 23 addresses within 3.4 hours into a single group, totaling 1,265 ETH. These transactions span Tornado Cash's 1, 10, and 100 ETH pools, with 80 deposits totaling 593 ETH occurring during the most active hour. The recovered group reveals a concentrated mixer-entry stage in which multiple addresses rapidly deposit funds across different denominations, indicating tightly coordinated execution across addresses and pools.

\noindent\textbf{KuCoin: cross-address, multi-day Mixer Exit.}
In the KuCoin case, \textsc{MixGuard} groups 228 withdrawals made by 19 addresses over 89.5 hours into a single group, totaling 22,800 ETH. All transactions are withdrawals from Tornado Cash's 100 ETH pool and are executed in multiple batches over nearly four days. The recovered group reveals a sustained, fixed-denomination mixer-exit stage spanning multiple addresses and time windows, thereby connecting separate withdrawal batches into one operation.

\noindent\textbf{Garden Finance: Recurring Railgun Deposit--Withdrawal Pattern.}
In the Garden Finance case, \textsc{MixGuard} groups 15 Railgun deposits and 7 withdrawals made by 8 addresses over approximately 116 hours. In one round, 6 addresses deposit within 2 hours, followed about 1 hour later by withdrawals from 4 of them within a 29-minute window. After 3 days, another address repeats the sequence within 75 minutes. The recovered group thus reveals a recurring deposit--delay--withdrawal pattern across days.


\subsection{Discovering Mixer Laundering in the Wild}
\label{subsec:wild}
\begin{figure}[t]
  \centering
  \begin{subfigure}[b]{0.38\linewidth}
    \centering
    \includegraphics[width=\linewidth]{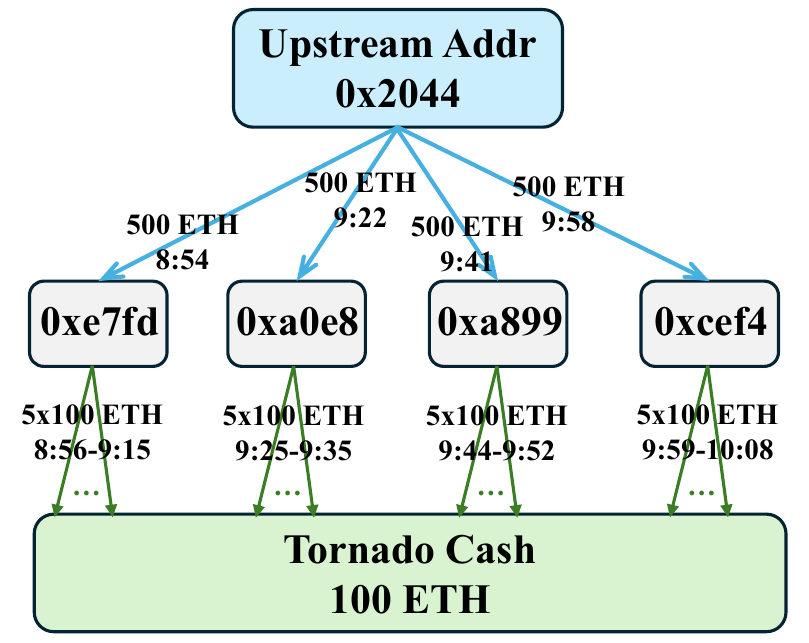}
    \caption{Saga case: co-funded peers.}
    \label{fig:saga_laundering}
  \end{subfigure}
  \begin{subfigure}[b]{0.38\linewidth}
    \centering
    \includegraphics[width=0.40\linewidth]{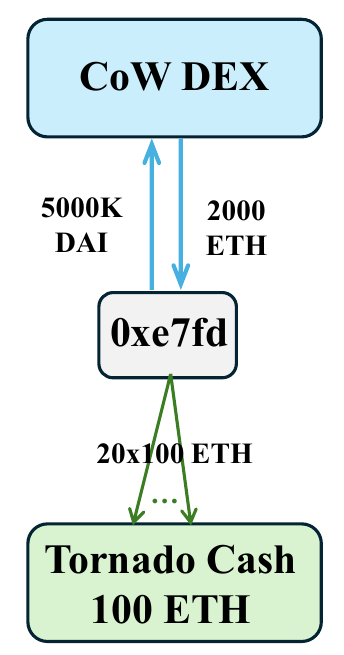}
    \caption{Coinbase: swap then deposit.}
    \label{fig:coinbase_laundering}
  \end{subfigure}
  \caption{Laundering patterns recovered by \textsc{MixGuard}.}
  \label{fig:wild_laundering}
\end{figure}




We deploy \textsc{MixGuard} \modified{in April 2026}, trained on \textsc{MixLaunder}, to real on-chain data from January 2026 to detect suspicious mixer laundering transactions. 
By cross-validating the detected results against contemporaneous incidents disclosed in security media, we find that \textsc{MixGuard} correctly identifies four real-world incidents involving over \textbf{\$15 million} in laundered funds. 
\modified{Of the 80 transactions flagged by \textsc{MixGuard},
55 were linked to these four incidents. Manual inspection of
the remaining 25 transactions found no clear signal of
laundering.}
As detailed in Table~\ref{tab:wild_cases}, these incidents differ substantially in attack type and upstream topology, indicating that \textsc{MixGuard} can identify previously unseen laundering incidents with different attack types and upstream flow patterns. 
\modified{As all four public reports preceded our analysis,
these results constitute retrospective detection rather than
prospective early warning.}
The deployment is lightweight in practice: a complete training run on \textsc{MixLaunder} takes 126 seconds on average, and inference scales near-linearly with the number of transactions. The detailed overhead results are reported in Appendix~\ref{subsec:overhead}.


Public reports tend to present such incidents in a fragmentary manner. One report may only identify the direct attacker without mentioning the downstream mixer laundering activities, while another may briefly describe the \emph{fund flow into Tornado Cash} without details. In contrast, \textsc{MixGuard} directly outputs an \textbf{observable, transaction-level} laundering graph. 
For example, Figure~\ref{fig:saga_laundering} shows how \textsc{MixGuard} successfully captures a coordinated laundering pattern: \emph{upstream funding $\rightarrow$ multi-wallet distribution $\rightarrow$ short-time parallel deposits into mixer}.
Figure~\ref{fig:coinbase_laundering} presents a different pattern of \emph{DEX swap $\rightarrow$ deposit into mixer} detected by \textsc{MixGuard}.
Furthermore, none of the suspicious addresses identified by \textsc{MixGuard} are flagged as suspicious by Ethereum explorers, e.g., Etherscan, showing that \textsc{MixGuard} surfaces stealthy laundering activity not yet covered by community monitoring. {We have responsibly disclosed our findings to Etherscan and Tornado Cash.}

\section{Discussion}
\label{sec-discussion}

\subsection{Implications}
Our study shows that although mixers conceal exact deposit--withdrawal links, laundering activities still leave observable behavioral and fund-flow evidence on both sides of the mixer. Based on these findings, we summarize some suggestions for  on-chain investigators, mixer developers, and blockchain intelligence providers.

\noindent\textbf{For on-chain investigators.} Existing deanonymization heuristics cover only a few cases, so investigators should not require an exact match between deposits and withdrawals as a prerequisite for identifying mixer laundering. 
Instead, they should combine address behavior, observable fund flows surrounding the mixer, and relationships among suspicious transactions to identify laundering activity and recover the corresponding mixer-entry or mixer-exit stages.

\noindent\textbf{For mixer developers.} Privacy protection need not preclude protocol-level risk response. Our analysis shows that absent or deposit-only screening may fail to flag suspicious activity. 
Developers should monitor observable deposits, withdrawals, and relayer interactions, issue auditable alerts for anomalous activity without exposing user identities or exact links, and update risk assessments for both current and historical transactions on both sides of the mixer.

\noindent\textbf{For blockchain intelligence providers.} The suspicious addresses identified by \textsc{MixGuard} during deployment were not flagged by Etherscan, indicating that existing intelligence data did not cover these addresses. 
Providers should complement address labels with transaction-level risk scores, transaction roles, evidence sources, and group-level associations to capture coordinated mixer activity spanning multiple addresses, transactions, and time windows.

\subsection{Limitations}

Our work carries certain limitations. 
First, for cases involving cross-chain transfers, we observe the Ethereum-side bridge interactions but do not yet connect them with fund flows on other chains. Our methods for transaction standardization and fund-flow reconstruction could be extended by treating bridges as edges between source and destination chains. 
Second, \textsc{MixLaunder} is constructed from publicly available evidence and on-chain records and therefore cannot include cases that have not yet been disclosed. \modified{Some laundering transactions within the included cases may
also remain unlabeled.}
\modified{
\textsc{MixLaunder} covers diverse cases, but large incidents account for most transactions, and cases involving \TC,
private-key compromises are overrepresented. Its behavioral statistics should not be interpreted as the overall distribution of mixer laundering activity.
}
However, \textsc{MixLaunder} can be continually expanded with newly disclosed cases and \textsc{MixGuard} detections that are subsequently confirmed as laundering cases through manual investigation.
Third, mixer privacy mechanisms conceal exact deposit--withdrawal correspondences, while private transfers within \RG are not observable on chain. Despite this boundary, our measurement and detection methods rely only on observable evidence on both sides of mixers, enabling laundering-transaction detection and case-aware grouping without compromising mixer privacy.

\section{Related Work}
\label{sec-rw}

\noindent\textbf{Mixer privacy and deanonymization.}
Prior work studies mixer design, usage, and privacy leakage. Glaeser et al.~\cite{glaeser2022foundations} formalize coin-mixing security, and Youn et al.~\cite{youn2023empirical} empirically characterize Tornado Cash usage. Wang et al.~\cite{wang2023zkmixer} show that ZKP mixers can be tied to DeFi attacks and that effective anonymity may be smaller than advertised. Linkage studies further infer deposit--withdrawal relations through learned models or heuristics, including MixBroker~\cite{du2024breaking}, MixLinker~\cite{Wang2025improving}, and Tornado Cash clustering~\cite{tc-clustering}. These works ask who corresponds to whom across the mixer boundary. We instead identify laundering-related mixer transactions and same-case groups without requiring explicit deposit--withdrawal linkage.

\vspace{2pt}
\noindent\textbf{Blockchain AML.}
Blockchain AML covers illicit classification, laundering-subgraph mining, group discovery, and fund-flow tracing. Weber et al.~\cite{weber2019anti} formulate AML as graph-based illicit transaction classification, while Elmougy et al.~\cite{elmougy2023demystifying} improve illicit-node detection through richer graph representations. DenseFlow~\cite{lin2024denseflow} mines laundering subgraphs, and MG-HRL~\cite{li2025multi} discovers organized laundering groups. \modified{FlowShield~\cite{fu2026flowshield} combines transaction semantics parsing and fund-flow reconstruction for transaction-level laundering detection.} However, these studies primarily rely on the ability to continuously observe fund flows through transaction relationships, and do not specifically address mixer laundering.

\vspace{2pt}
\noindent\textbf{Ethereum incident analysis.}
Incident-grounded Ethereum studies reconstruct illicit behavior from real-world cases. EthereumHeist~\cite{wu2023towards} characterizes asset flows originating from Ethereum heists, while Fu et al.~\cite{fu2023bigger} analyze the laundering gang associated with the Upbit hack. MFTracer~\cite{huo2025shedding} automates fine-grained transaction-level tracing of illicit flows on EVM-compatible blockchains. Related transaction-centric studies characterize and detect specific threats, including transaction-based phishing~\cite{he2023txphishscope} and scam tokens~\cite{wu2024tokenscout}. We follow this incident-grounded perspective but focus on mixer laundering, where the privacy boundary interrupts observable fund-flow continuity. We therefore identify laundering-related mixer transactions and same-case groups from evidence on both sides of the mixer without requiring exact deposit--withdrawal links.

\section{Conclusion}
\label{sec-conclu}
This paper presents the first comprehensive measurement study of mixer laundering on Ethereum. We begin by building \textsc{MixLaunder}, a public case-level dataset covering 27 real-world laundering cases involving \TC and \RG, with labeled mixer deposits, withdrawals, and upstream/downstream fund flows.
Using this dataset, we measure the full laundering workflow before, through, and after mixers, and summarize five common strategies that distinguish laundering from background mixer usage. Our analysis also shows why existing mixer defenses and deanonymization heuristics provide limited coverage in real laundering cases.
Guided by these measured strategies, we design \textsc{MixGuard}, which supports suspicious mixer-transaction detection and case-aware group recovery.
Our dataset, measurements, and detector provide a practical foundation for understanding and mitigating mixer laundering without breaking mixer privacy. 

\section*{Acknowledgments}
We would like to thank the anonymous reviewers and shepherd for their comments that greatly helped improve the presentation of this paper. This work is partially supported by the Australian Research Council through Discovery Early Career Researcher Award (DE250100032).

\bibliographystyle{ACM-Reference-Format}
\bibliography{bib}

\appendix

\section{Additional Details on Dataset Construction}
\label{app:dataset}

\subsection{Public Clue for Mixer Laundering Cases}
\label{app:case-clues}

Public investigation materials provide two types of clues for mixer laundering
cases.

\vspace{5pt}
\noindent\textbf{Deposit-side clues.}
Deposit-side clues are usually direct because the movement of stolen funds into
a mixer is publicly visible. These clues may include the transaction hash of a
mixer deposit, the address that sends funds to the mixer, the mixer pool or
relay involved in the deposit, or a fund-flow graph showing stolen funds entering
the mixer. \modified{
For example, the CoinMetro investigation~\cite{zachxbt_lazarus_laundered_200m} describes how stolen funds
were transferred through intermediate addresses and consolidated before
being deposited into \TC. These pre-mixing transfers provide on-chain
evidence supporting the report's association of the deposits with
case-related funds.
}

\vspace{5pt}
\noindent\textbf{Withdrawal-side clues.}
Withdrawal-side clues are usually reconstructed from observable on-chain
signals. Public investigators do not claim to break the cryptographic link
between a specific deposit and a specific withdrawal. Instead, they first collect
candidate withdrawal events within a relevant time window, and then filter or
group them using multiple on-chain signals. These signals may include asset
type, amount range, withdrawal batch, gas funding source, post-withdrawal
bridging, aggregation behavior, flow to known exchanges or P2P platforms, and
other service interactions. \modified{
For example, the CoinMetro investigation~\cite{zachxbt_lazarus_laundered_200m} examined withdrawals during
the relevant period and identified a batch sent to a single address.
The report also noted a fund-flow connection between this withdrawal
address and the original theft address, providing further evidence
linking the withdrawal address to the case.
}


\modified{
We use the case attributions and transaction clues provided by these investigations to collect and check the corresponding on-chain mixer transactions, as described in \S\ref{sub-case}.
}

\subsection{Mixer Transaction Extraction Rules}
\label{app:mixer-extraction}

This part explains how we extract a mixer deposit or withdrawal
record from an on-chain transaction. A mixer-related transaction may contain
multiple fund transfers, including user fund transfers, internal contract
routing, relayer fees, gas-related payments, wrap or unwrap operations, and
other intermediate contract calls. To avoid counting auxiliary transfers as
independent mixer deposit or withdrawal records, we only keep the core transfer
that represents user funds entering or leaving the mixer.

\noindent\textbf{Deposit extraction.}
For a deposit transaction, we choose the transfer where funds move from the
user-side address into a mixer-related contract as the core transfer. For \TC,
we keep the incoming transfer from the user-side address to a TC pool, proxy, or
router. If the transaction later contains an internal transfer from the proxy or
router to the pool, we use that internal transfer only to confirm that funds
entered the pool. We do not keep it as the standardized deposit record. For
\RG, the core transfer usually moves user funds into the privacy system through
a relay or helper contract. If the transaction also contains fee payments,
router-internal transfers, or intermediate contract calls, we do not treat these
auxiliary transfers as independent deposit records.


\vspace{5pt}
\noindent\textbf{Withdrawal extraction.}
For a withdrawal transaction, we choose the transfer where funds are released
from a mixer-related contract to the withdrawal address as the core transfer.
For \TC, this usually means the pool releases a fixed-denomination amount to the
recipient address. For \RG, this usually means the privacy system releases funds
to a public address through a relay or helper contract. We only keep the main
transfer to the withdrawal address. We do not treat relayer fees, internal
routing, or auxiliary contract transfers as independent withdrawal records.

\newcommand{\algcomment}[1]{\hspace{1.2em}$\rhd$ \textit{#1}}

\begin{algorithm}[t]
\caption{Fund-flow tracing}
\label{alg:fund-flow-tracing}
\begin{algorithmic}[1]
\STATE \textbf{Input:} Mixer seed records $\mathcal{S}$, direction $d \in \{\textsf{up}, \textsf{down}\}$
\STATE \textbf{Output:} Attributed fund-flow records $\mathcal{E}$
\STATE Initialize tracing queue $\mathcal{Q}$ as empty \algcomment{\textcolor{blue}{stores states to trace}}
\FORALL{seed $s \in \mathcal{S}$}
    \STATE $a_0 \leftarrow \mathrm{StartAddress}(s,d)$ \algcomment{\textcolor{blue}{deposit sender or withdrawal receiver}}
    \STATE $q \leftarrow (a_0, s.\mathrm{asset}, s.\mathrm{amount}, s.\mathrm{time}, 0, s.\mathrm{tx})$
        \algcomment{\textcolor{blue}{initial state}}
    \STATE Push $q$ into $\mathcal{Q}$ \algcomment{\textcolor{blue}{start tracing this seed}}
\ENDFOR
\WHILE{$\mathcal{Q}$ is not empty}
    \STATE Pop $q=(a,x,v,t,h,sid)$ from $\mathcal{Q}$ \algcomment{\textcolor{blue}{current address, asset, amount}}
    \STATE $\ell \leftarrow \mathrm{QueryLabel}(a)$ \algcomment{\textcolor{blue}{check service label}}
    \IF{$v=0$ or $\ell$ is a terminal service}
        \STATE \textbf{continue} \algcomment{\textcolor{blue}{stop this path}}
    \ENDIF
    \STATE $\mathcal{T} \leftarrow \mathrm{Candidates}(a,d)$ \algcomment{\textcolor{blue}{select transfers by direction}}
    \FORALL{$\tau \in \mathcal{T}$}
        \IF{$\tau$ does not match asset $x$ or time order $t$}
            \STATE \textbf{continue} \algcomment{\textcolor{blue}{skip unrelated transaction}}
        \ENDIF
        \STATE $v' \leftarrow \mathrm{AttributeAmount}(q,\tau)$ \algcomment{\textcolor{blue}{FIFO attribution}}
        \IF{$v'=0$}
            \STATE \textbf{continue} \algcomment{\textcolor{blue}{no attributed amount}}
        \ENDIF
        \IF{$\tau$ is a swap}
            \STATE $(a',x',v') \leftarrow \mathrm{ResolveSwap}(\tau,d,v')$
                \algcomment{\textcolor{blue}{map input/output assets}}
        \ELSE
            \STATE $(a',x') \leftarrow \mathrm{NextState}(\tau,d)$ \algcomment{\textcolor{blue}{next hop}}
        \ENDIF
        \STATE Record the attributed fund-flow edge in $\mathcal{E}$ \algcomment{\textcolor{blue}{save traced edge}}
        \STATE Push $(a',x',v',\tau.\mathrm{time},h+1,sid)$ into $\mathcal{Q}$
            \algcomment{\textcolor{blue}{continue from next hop}}
    \ENDFOR
\ENDWHILE
\RETURN $\mathcal{E}$
\end{algorithmic}
\end{algorithm}

\subsection{Fund-Flow Tracing Details}
\label{app:tracing}

This section gives the detailed rules for fund-flow tracing and the
corresponding pseudocode in Algorithm~\ref{alg:fund-flow-tracing}. The input is
a set of standardized mixer seed records and a tracing direction
$d \in \{\textsf{up}, \textsf{down}\}$. The output is a set of attributed
fund-flow records.


\vspace{5pt}
\noindent\textbf{Tracing state.}
For each mixer seed, we maintain a queue of tracing states. A state contains the
current address, asset type, attributable amount, timestamp, hop count, and seed
transaction hash. The seed transaction hash stays unchanged during tracing and
identifies which mixer transaction the traced path belongs to. The current
address, asset, and amount are updated as funds are transferred, split, or
swapped. For upstream tracing, the start address is the sender of the mixer
deposit. For downstream tracing, the start address is the receiver of the mixer
withdrawal.


\vspace{5pt}
\noindent\textbf{Candidate selection.}
At each hop, we query the label of the current address and collect candidate
transactions according to the tracing direction. For downstream tracing, we
select outgoing transactions after the funds arrive at the current address. For
upstream tracing, we select incoming transactions before the mixer deposit that
may explain the deposit source. A candidate transaction must match the current
asset type and satisfy the required time order.

\vspace{5pt}
\noindent\textbf{Amount attribution.}
For each candidate transaction, we compute how much of its transferred amount
can be attributed to the current tracing state. We use a FIFO-style attribution
rule: funds that arrive earlier at an address are consumed before funds that
arrive later. If no amount can be attributed, the candidate transaction is
discarded. Otherwise, we record only the attributed amount and continue tracing
from the next address.

\vspace{5pt}
\noindent\textbf{Swap handling.}
If a candidate transaction involves a swap, we parse event logs and internal
transfers to connect the input asset and output asset. For downstream tracing,
we follow the output asset after the swap. For upstream tracing, we trace back
to the input asset before the swap. This allows the traced path to continue when
funds change assets.


\vspace{5pt}
\noindent\textbf{Stopping conditions.}
Tracing stops for a path when the attributable amount becomes zero, the current
address is labeled as a terminal service, no candidate transaction passes the
asset and time filters, or no candidate transaction receives a positive
attributed amount. Terminal services are identified by querying address labels
and include exchanges, bridges, and other labeled service addresses. Each output
record contains the seed transaction hash, current transaction hash, sender
address, receiver address, asset type, attributed amount, timestamp, and hop
count.

\section{Hyperparameter Investigation}
\label{subsec:hyper}

To evaluate hyperparameter sensitivity, we vary the encoder loss weight $\alpha$ and the clustering parameters $k$ and $\tau_s$. We report their effects on the four main metrics under the same 9-fold cross-validation setting as the main results.

\begin{figure}[t]
\centering

\begin{subfigure}[t]{0.21\linewidth}
    \centering
    \includegraphics[width=\linewidth]{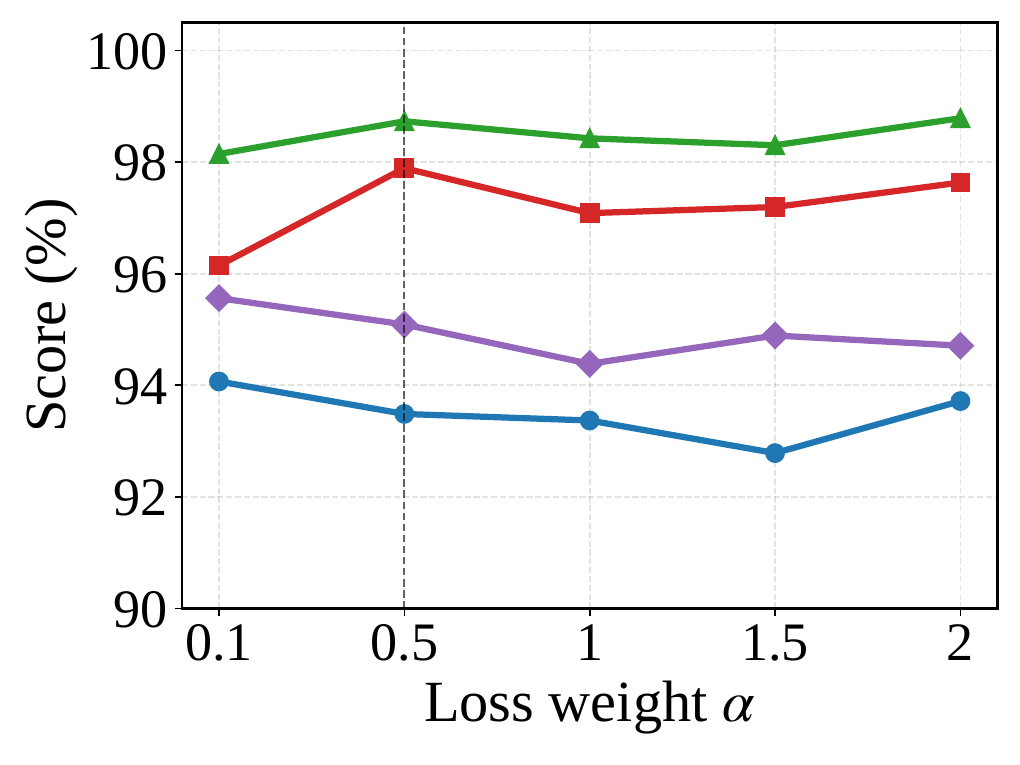}
    \caption{Loss weight $\alpha$.}
    \label{fig:alpha_sensitivity}
\end{subfigure}
\begin{subfigure}[t]{0.76\linewidth}
    \centering
    \includegraphics[width=\linewidth]{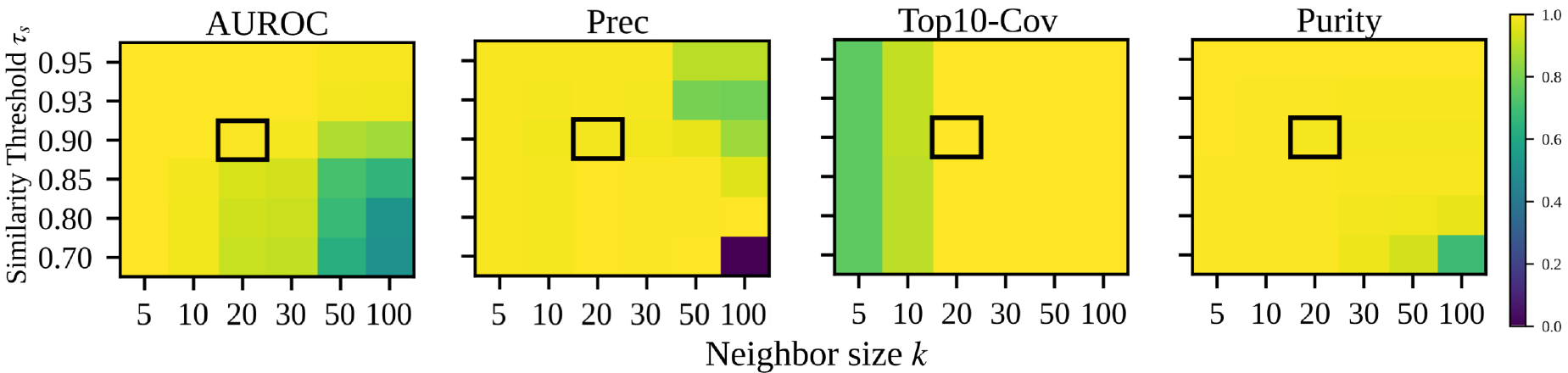}
    \caption{Neighbor size $k$ and similarity threshold $\tau_s$.}
    \label{fig:sensitivity_stage2}
\end{subfigure}

\caption{Hyperparameter sensitivity of \textsc{MixGuard}. 
(a) Sensitivity to the encoder loss weight $\alpha$. 
(b) Sensitivity to the grouping parameters $k$ and $\tau_s$ across four metrics.
The black box marks the default setting $(k{=}20,\tau_s{=}0.90)$.}
\label{fig:hyper_sensitivity}

\end{figure}

\smallskip
\noindent\emph{Loss weight $\alpha$.}
Figure~\ref{fig:alpha_sensitivity} sweeps $\alpha\in\{0.1,0.5,1.0,1.5,2.0\}$,
covering a $20\times$ range that spans BCE-dominant and SupCon-dominant
regimes. \textbf{All four metrics stay nearly flat across the full
range}: \texttt{Purity} varies by only $0.65$ percentage points
($98.14\%$--$98.79\%$) with per-fold std below $1.05\%$, and
\texttt{AUROC}, \texttt{Prec}, and \texttt{Top10-Cov} all remain
within their per-fold std of one another. The default
$\alpha{=}0.5$ therefore sits comfortably inside a stable plateau and
is not the result of careful tuning.

\smallskip
\noindent\emph{Neighbors size $k$ and similarity threshold $\tau_s$.}
Figure~\ref{fig:sensitivity_stage2} shows that \textbf{\textsc{MixGuard}
is robust to the choice of $k$ and $\tau_s$, with all four metrics
staying close to their best values across most of the grid.}
Visible variation appears mainly near the boundary of the search
space. When $k$ is very small, such as $5$, the mutual $k$-NN graph
becomes sparser, which splits transactions from the same case into
several small clusters and lowers \texttt{Top10-Cov} to $0.75$,
while the other three metrics remain close to their best values.
When $k$ is large and $\tau_s$ is low, such as $k{=}100$ and
$\tau_s{=}0.70$, the graph becomes more connected and leads to more
merging, which reduces \texttt{AUROC}, \texttt{Prec}, and
\texttt{Purity}. Based on these results, we use $k{=}20$ and
$\tau_s{=}0.90$ as the default setting, which lies \textbf{near the
center of the stable region and maintains strong performance under
small changes} in data distribution or graph structure.



\begin{table}[t]
\centering
\caption{Summary of observable feature groups used by \textsc{MixGuard}.}
\label{tab:mixguard-features}
\small
\setlength{\tabcolsep}{4pt}
\renewcommand{\arraystretch}{1.08}
\begin{tabular}{
>{\raggedright\arraybackslash}p{0.27\columnwidth}
>{\raggedright\arraybackslash}p{0.65\columnwidth}}
\toprule
\textbf{View / Module} & \textbf{Feature Summary} \\
\midrule
Address view
& Actor role, mixer-use frequency, address reuse, activity, transaction timing, counterparties, and service interactions. \\

Path view
& Path duration, transaction rate and timing, fan-in/fan-out, and observed fund-flow behaviors. \\

Topology view
& Local fund-flow connectivity, node degrees, hop positions, and actor-node information. \\

Weak-tie merger
& Cross-cluster edge statistics, embedding and structural similarity, service consistency, cluster scale, and temporal alignment. \\
\bottomrule
\end{tabular}
\end{table}

\begin{figure}[]
  \centering
  \begin{subfigure}[b]{0.4\linewidth}
    \centering
    \includegraphics[width=\linewidth]{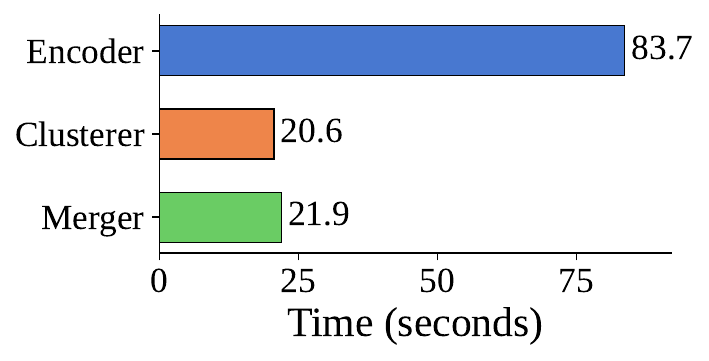}
    \caption{Per-stage training time.}
    \label{fig:overhead}
  \end{subfigure}
       \hspace{0.02\linewidth}
  \begin{subfigure}[b]{0.4\linewidth}
    \centering
    \includegraphics[width=\linewidth]{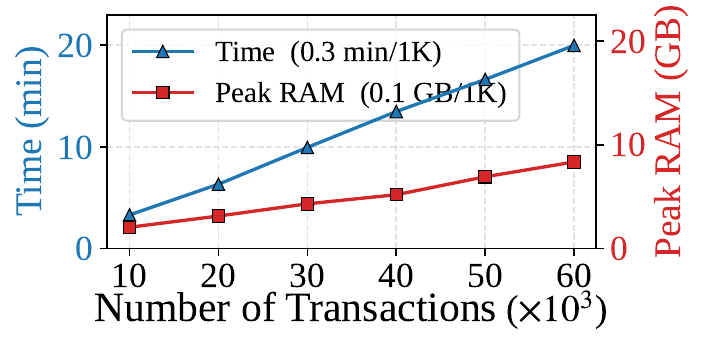}
    \caption{Deployment scaling.}
    \label{fig:scaling}
  \end{subfigure}
  \caption{System overhead of \textsc{MixGuard}. 
  (a) Per-stage time during training on \textsc{MixLaunder}. 
  (b) Inference time and peak RAM during deployment}
  \label{fig:system_overhead}
\end{figure}

\section{Experimental Configuration Details}
\label{app:exp-config}

\noindent\textbf{Feature groups.}
Table~\ref{tab:mixguard-features} summarizes the feature groups used by \textsc{MixGuard}.

\noindent\textbf{Hardware and software.}
We implement \textsc{MixGuard} in PyTorch~2.4.1 (CUDA~12.4) and PyTorch Geometric~2.7. All experiments run on Ubuntu~22.04 with a single NVIDIA~RTX~4090 GPU (24~GB VRAM), an AMD~Ryzen Threadripper~7960X CPU (24~cores / 48 threads), and 256~GB RAM under Python~3.11. The random seed is fixed to~17 for all nine folds.

\noindent\textbf{Tri-view encoder.}
The address and path MLPs and the two-layer GraphSAGE share hidden dimension $d{=}128$ with Dropout~$0.1$. We optimize with AdamW (lr~$10^{-3}$, weight decay~$10^{-4}$). Each batch is drawn by a case-balanced sampler with $P{=}4$ cases, $K{=}8$ laundering transactions per case, and $B{=}16$ background transactions. The total loss combines BCE detection and masked supervised contrastive loss with $\alpha{=}0.5$ and temperature $\tau{=}0.1$.

\noindent\textbf{Grouping modules.}
For the strong-tie clusterer, we build a mutual top-$k$ nearest-neighbor graph on the $\ell_2$-normalized projections with $k{=}20$ and similarity threshold $\tau_s{=}0.90$. For the weak-tie merger, the merge classifier is LightGBM with 400 trees, learning rate $0.03$, 31 leaves, and subsample rate $0.9$.

\noindent\textbf{\modified{Baseline configurations.}}
\modified{K-means and HDBSCAN are applied to the concatenated address and path features after $\ell_2$ normalization. K-means uses $k{=}3$ and 10 initializations, while HDBSCAN uses a minimum cluster size of 5 and the EOM cluster-selection method. The MLP and LightGBM baselines use the same address and path features for binary detection, with predicted laundering probabilities
thresholded at $0.5$; their detected transactions are subsequently grouped using HDBSCAN. LightGBM uses 400 trees, a learning rate of $0.05$, and 31 leaves. For DenseFlow, MG-HRL, MixBroker, and MixLinker, we retain their original representations and adapt their predicted subgraphs or links into transaction groups for evaluation under the same case-level splits.}

\begin{table*}[t]
\centering
\caption{\modified{Nine-fold results of MixGuard. SD denotes the standard deviation across the nine folds.}}
\label{tab:mixguard-nine-fold}
\setlength{\tabcolsep}{6pt}
\begin{tabular}{lrrrrrrr}
\toprule
Fold & AUROC(\%) & AP(\%)  & Precision(\%)  & Recall(\%)  & F1(\%)  & Purity(\%)  & Top10-Cov(\%)  \\
\midrule
0 & 99.89 & 99.46 & 97.62 & 99.19 & 98.40 & 98.87 & 100.00 \\
1 & 98.61 & 96.32 & 99.36 & 85.25 & 91.76 & 96.33 & 100.00 \\
2 & 97.99 & 95.38 & 91.77 & 92.62 & 92.19 & 97.58 & 93.96 \\
3 & 91.50 & 83.84 & 100.00 & 69.97 & 82.33 & 98.45 & 96.80 \\
4 & 89.48 & 79.94 & 99.69 & 66.22 & 79.58 & 98.51 & 88.93 \\
5 & 89.73 & 76.67 & 94.32 & 54.87 & 69.38 & 98.83 & 99.94 \\
6 & 95.98 & 87.05 & 80.76 & 91.14 & 85.64 & 98.48 & 97.61 \\
7 & 96.10 & 92.61 & 98.96 & 87.76 & 93.03 & 99.24 & 86.54 \\
8 & 85.04 & 69.50 & 99.38 & 48.37 & 65.07 & 99.29 & 99.81 \\
\midrule
\textbf{SD}
  & 5.06 & 10.10 & 6.29 & 18.01 & 11.25 & 0.93 & 5.11 \\
\bottomrule
\end{tabular}
\end{table*}

\begin{table*}[t]
\centering
\caption{\modified{Performance by evaluated case size.}}
\label{tab:mixguard-case-size}
\small
\setlength{\tabcolsep}{4pt}
\begin{tabular}{lrrrrrrrr}
\toprule
Case size & Cases & Laundering Tx & AUROC(\%) & AP(\%) & Precision(\%) & Recall(\%) & F1(\%) & Top10-Cov(\%) \\
\midrule
Small  & 9 & 272 & 90.88 & 63.78 & 69.47 & 62.25 & 62.88 & 100.00 \\
Medium  & 9 & 897 & 89.74 & 60.78 & 68.05 & 68.29 & 62.31 & 98.84 \\
Large  & 9 & 8131 & 92.86 & 82.45 & 94.92 & 73.79 & 81.40 & 89.03 \\
\bottomrule
\end{tabular}
\end{table*}

\section{Additional Evaluation Results}
\label{subsec:A-results}

\modified{Table~\ref{tab:mixguard-nine-fold} reports results for each of the nine folds and the standard deviations across folds, including additional metrics of average precision (AP), recall, and F1. 
We categorize cases by their number of labeled
laundering transactions into small ($<50$), medium
($50$--$149$), and large ($\geq150$) groups.
Table~\ref{tab:mixguard-case-size} reports performance for
these three groups, with nine cases in each group.} \modified{For this analysis, we compute each metric separately for every held-out case and report the unweighted mean across the nine cases in each size group. Thus, each case contributes equally regardless of its number of laundering transactions.}

\section{System Overhead}
\label{subsec:overhead}



\noindent{\textbf{Training Overhead.}} 
\textit{Training \textsc{MixGuard} is lightweight.} Figure~\ref{fig:overhead} shows that a complete training run on the \textsc{MixLaunder} dataset takes only 126 seconds on average. \emph{Tri-view encoder} training dominates the cost at 66.3\% of the total, while \emph{strong-tie clusterer} and \emph{weak-tie merger} account for the remaining 33.7\%, each averaging less than 22 seconds.


\smallskip
\noindent{\textbf{Deployment Overhead.}}
\textsc{MixGuard} maintains \textit{scalable resource overhead under large-scale deployment}. To assess scalability, we synthesize test sets of varying sizes via bootstrap sampling with Gaussian noise, and measure the resource cost of the full inference pipeline at each size. As shown in Figure~\ref{fig:scaling}, when the transaction count grows from 10K to 60K, inference time scales from $\sim$3 to $\sim$20 minutes and peak memory usage from 2.1~GB to 8.5~GB, both exhibiting near-linear behavior.

\section{Example Mixer Laundering Report}
\label{app:case-report}






To support manual inspection and make the traced fund flows easier to audit, we
prepare a mixer laundering report for each incident in \textsc{MixLaunder}.
Each report summarizes the structured tracing results of one incident, including
the mixer deposits and withdrawals, the involved mixer pools or contracts, key
addresses, major upstream sources, major downstream destinations, and the main
fund-flow patterns. The goal of the report is not to add extra off-chain
identity information, but to present the on-chain tracing results in a compact
and readable form.

\begin{figure}[t]
    \centering
    \includegraphics[width=0.85\columnwidth]{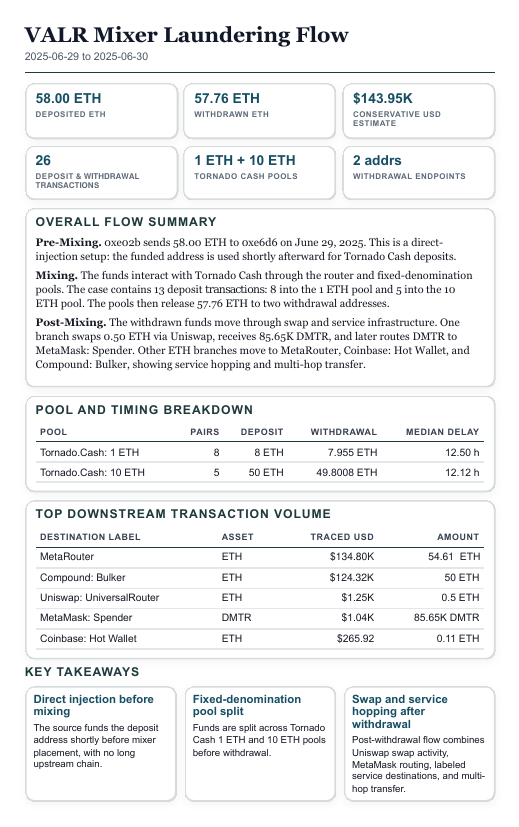}
    \caption{Case summary for the VALR incident.}
    \label{fig:valr-report-summary}
\end{figure}

\begin{figure}[t]
    \centering
    \includegraphics[width=0.85\columnwidth]{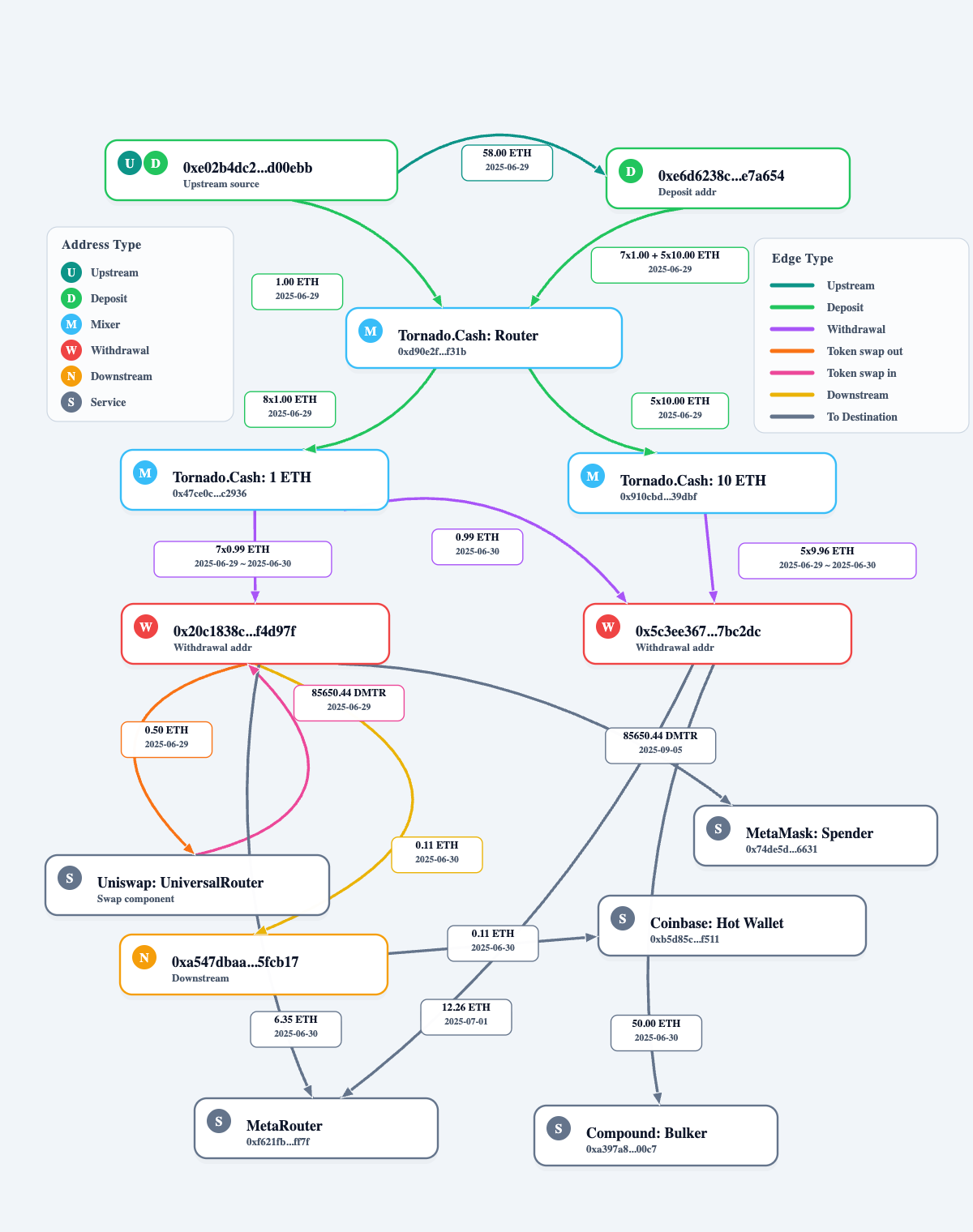}
    \caption{Fund-flow visualization for the VALR incident.}
    \label{fig:valr-report-flow}
\end{figure}

Figure~\ref{fig:valr-report-summary} and Figure~\ref{fig:valr-report-flow} show the report for the VALR incident as an
example. The report contains two parts. The first part is a case summary~(Figure~\ref{fig:valr-report-summary}),
covering the number of deposits and withdrawals, the Tornado Cash pools used,
the matched mixer transactions, and the main downstream destinations. The second
part is a fund-flow visualization~(Figure~\ref{fig:valr-report-flow}), showing how funds move from upstream
addresses, through the Tornado Cash router and the 1 ETH and 10 ETH pools, to
withdrawal addresses and downstream services. We release the reports
for all cases together with our dataset and code.





\section{Ethical Considerations}

This work studies mixer laundering using public blockchain data, public incident reports, and protocol documentation. We do not contact victims, attackers, or mixer operators during data collection, and we do not actively intervene in live systems. Our study does not involve human subjects, private user accounts, or non-public transaction data.

The main ethical risk is dual use. Our analysis can help defenders understand how illicit funds are prepared, mixed, and redeployed, but it may also reveal investigative blind spots. We reduce this risk in three ways. First, the paper focuses on defensive goals: detecting laundering-related mixer transactions and recovering same-case groups, rather than improving attacker privacy or evasion. Second, our analysis uses historical cases and public on-chain behavior, and does not disclose private information, credentials, or non-public operational details. Third, for new suspicious activity identified in our 2026 deployment study, we follow responsible disclosure and report findings to Etherscan and Tornado Cash before broad dissemination.

We believe the benefits outweigh the risks. Mixer laundering has become an important part of Ethereum post-attack cash-out workflows, while current defenses remain limited. By releasing a case-grounded dataset, identifying failure modes of existing defenses, and proposing a detection framework, our work helps investigators, analytics platforms, and protocol operators respond to real-world laundering activity.

\section{Generative AI Usage}



We used a generative AI tool for language editing and feedback on the organization and limited revision of non-technical prose. The research design, data collection, implementation, experiments, analysis, and conclusions were conducted by the authors. All AI-assisted text was reviewed and revised by us.

\received{10 July 2026}
\received[revised]{5 October 2026}
\received[accepted]{9 October 2026}

\end{document}